# Constraints for precise and accurate fluid inclusion stable isotope analysis using water-vapour saturated CRDS techniques

Therese Weissbach[1,2], Tobias Kluge[1,2,3,4,*], Stéphane Affolter[5], Markus C. Leuenberger[6], Hubert Vonhof[7], Dana F.C. Riechelmann[8], Jens Fohlmeister[9,10], Marie-Christin Juhl[2], Benedikt Hemmer[2], Yao Wu[2], Sophie F. Warken[2,11], Martina Schmidt[2], Norbert Frank[2], Werner Aeschbach[2,3]

[1]Heidelberg Graduate School of Fundamental Physics, Heidelberg University, Im Neuenheimer Feld 226, 69120 Heidelberg, Germany

[2]Institute of Environmental Physics, Heidelberg University, Im Neuenheimer Feld 229, 69120 Heidelberg, Germany

[3]Heidelberg Center for the Environment, Heidelberg University, Im Neuenheimer Feld 229, 69120 Heidelberg, Germany

[4]now at: Institute of Applied Geosciences, Karlsruhe Institute of Technology, Adenauerring 20b, 76131 Karlsruhe, Germany

[5]Department of Environmental Sciences, University of Basel, Bernoullistrassse 30/32, 4056 Basel, Switzerland

[6]Climate and Environmental Physics Division, Physics Institute and Oeschger Centre for Climate Change Research, University of Bern, Sidlerstrasse 5, 3012 Bern, Switzerland

[7]Climate Geochemistry Department, Max Planck Institute for Chemistry, Hahn-Meitner-Weg 1, 55128 Mainz, Germany

[8]Institute for Geosciences, Johannes Gutenberg University Mainz, Johann-Joachim-Becher-Weg 21, 55128 Mainz, Germany

[9]Federal Office for Radiation Protection, Köpenicker Allee 120-130, 10318 Berlin, Germany

[10]GFZ German Research Centre for Geosciences, Section 'Climate Dynamics and Landscape Development', Telegrafenberg, 14473 Potsdam, Germany

[11]Institute of Earth Sciences, Heidelberg University, Im Neuenheimer Feld 234, 69120 Heidelberg, Germany

*corresponding author: tobias.kluge@kit.edu

**Journal: Chemical Geology**

**Highlights:**

- Laser-based fluid inclusion analysis with water-vapour purged extraction enables high precision ($\delta^2H \leq \pm1.5‰$, $\delta^{18}O \leq \pm0.5‰$)
- Biasing effects (memory, adsorption, amount) in fluid inclusion isotope analysis are negligible for ≥1 μl water/g calcite
- Isotopic interference is negligible for sample isotope ratios within 10‰($\delta^{18}O$) and 50‰($\delta^2H$) of the water vapour background
- Reconstructed temperatures of a 20[th] century stalagmite trace the recent warming of 1 °C in Central Europe

**Abstract**

Hydrogen ($\delta^2$H) and oxygen ($\delta^{18}$O) isotopes of water extracted from speleothem fluid inclusions are important proxies used for paleoclimate reconstruction. In our study we use a cavity ring-down laser spectroscopy system for analysis and modified the approach of Affolter et al. (2014) for sample extraction. The method is based on crushing of small sub-gram speleothem samples in a heated and continuously water-vapour purged extraction line. The following points were identified:

Injection of reference water shows a precision (1σ) of 0.4-0.5 ‰ for $\delta^{18}$O values and 1.1-1.9 ‰ for $\delta^2$H values for water amounts of 0.1-0.5 µl, which improves with increasing water amount to 0.1-0.3 ‰ and 0.2-0.7 ‰, respectively, above 1 µl. The accuracy of measurements of water injections and water-filled glass capillaries crushed in the system is better than 0.08 ‰ for $\delta^{18}$O and 0.3 ‰ for $\delta^2$H values. The reproducibility (1σ) based on replicate analysis of speleothem fluid inclusion samples with water amounts > 0.2 µl is 0.5 ‰ for $\delta^{18}$O and 1.2 ‰ for $\delta^2$H values, respectively. Isotopic differences between the water vapour background of the extraction system and the fluid inclusions have no significant impact on the measured fluid inclusion isotope values if they are within 10 ‰ for $\delta^{18}$O and 50 ‰ for $\delta^2$H values of the background. Tests of potential adsorption effects with inclusion free spar calcite confirm that the isotope values are unaffected by adsorption for water contents of about 1 µl (fluid inclusion) water per g of carbonate or above.

Fluid inclusion analysis on three different modern to late Holocene speleothems from caves in northwest Germany resulted in $\delta^{18}$O and $\delta^2$H values that follow the relationship as defined by the meteoric water line and that correspond to the local drip water. Yet, due to potential isotope exchange reactions for oxygen atoms, hydrogen isotope measurements are preferentially to be used for temperature reconstructions. We demonstrate this in a case study with a Romanian stalagmite, for which we reconstruct the 20$^{th}$ century warming with an amplitude of approximately 1 °C, with a precision for each data point of better than ±0.5 °C.

## 1. *Introduction*

Speleothem fluid inclusions can provide direct insight into past climatic conditions as they are a unique archive for the original drip water and the corresponding meteoric water (e.g., Griffiths et al., 2010; Affolter et al., 2014; Labuhn et al., 2015; Warken et al., 2022). Fluid inclusion water isotope ratios ($\delta^{18}$O and $\delta^{2}$H values) are increasingly usedas proxies in hydrology and paleoclimate studies (e.g., McGarry et al., 2004; Demény et al., 2017; Millo et al., 2017; Affolter et al., 2019; Wilcox et al., 2020; Matthews et al., 2021). Two physically different measurement principles, laser spectroscopy (mainly cavity ring-down spectroscopy - CRDS) and isotope ratio mass spectrometry (IRMS), allow determining the isotopic composition of speleothem fluid inclusion water (CRDS: Arienzo et al., 2013; Affolter et al., 2014; Uemura et al., 2016; Dassié et al., 2018; IRMS: Dennis et al., 2001; Vonhof et al., 2006; Dublyansky and Spötl, 2009). Although CRDS and IRMS systems yield comparable results (de Graaf et al., 2020) challenges remain for both methods regarding precise and reproducible analysis of small water amounts. Often only a single measurement attempt is possible due to low growth rates of the speleothems (often 10-100 µm/a) or intended high resolution. Water contents in natural speleothems range from ~ 0 up to several 10 µl per g (McDermott et al., 2006). The necessary water sample amount (depending on the setup 0.05-0.2 µl, e.g., Dublyanksy and Spötl, 2009; Uemura et al., 2016) limits the temporal resolution and restricts analytical repetition.

Fluid inclusion water for isotope analysis is released either by crushing (e.g., Schwarcz et al., 1976; Dennis et al., 2001; Vonhof et al., 2006; Dublyansky and Spötl, 2009; Demény et al., 2013) or thermal decrepitation (e.g., Yonge, 1982; McGarry et al., 2004; Verheyden et al., 2008). Thermal decrepitation has the disadvantage that structurally-bound water with a very low $\delta^{2}$H value may be released during extraction, resulting in large isotopic shifts of up to 30 ‰ in comparison to parent cave drip water (Yonge, 1982; Matthews et al., 2000; McGarry et al., 2004; Verheyden et al., 2008). This analytical artefact can be largely avoided by crushing the sample mechanically. For fluid inclusion analysis using IRMS (Schwarcz et al., 1976; Harmon et al., 1978; 1979), water was extracted by crushing the sample under vacuum conditions and then subsequently converted to water vapour followed by conversion into directly measurable gases such as $H_2$ for H isotopic analysis. Fluid inclusion $\delta^{18}$O values were initially not measured but calculated from measured $\delta^{2}$H values via the relationship between $\delta^{18}$O and $\delta^{2}$H values of the meteoric water line (e.g., Schwarcz et al., 1976). In recent extraction systems the oxygen in fluid inclusion water is converted to CO gas during a high temperature reaction with glassy carbon which is then used for analysis of $\delta^{18}$O values in an IRMS (e.g., Dublyansky and Spötl, 2009). The first combined method for oxygen and hydrogen measurements with an off-line crushing method and dual-inlet IRMS was developed by Dennis et al. (2001). It achieved good precision of ± 0.4 ‰ for $\delta^{18}$O and ± 3 ‰ for $\delta^{2}$H values, but required a comparatively large sample with water amounts of 1-3 $\mu$l (see also Matthews et al., 2000). A reduction in sample amount down to 0.1 $\mu$l, which corresponds to 0.1 g of calcite for samples that contain 1 μl of water per gram, was achieved by Vonhof et al. (2006) by combining off-line preparation and continuous-flow mass spectrometry. This technique enables a faster analysis of 0.1 - 0.2 *μ*l sized samples with a precision of ± 0.5 ‰ for $δ^{18}O$ and ± 1.5 ‰ for $δ^2H$ values (Vonhof et al., 2006; Dublyansky and Spötl, 2009; de Graaf et al., 2020).

Laser spectroscopy is less expensive and represents a reliable, precise and easy technique to directly measure stable water isotopes (Brand et al., 2009; Gupta et al., 2009). The first application using CRDS to measure fluid inclusions in speleothems was developed by Arienzo et al. (2013). They used a CRDS analyser with a stainless-steel line heated to 115 °C that was constantly flushed with dry nitrogen as a carrier gas. It achieved comparable precisions as the traditional IRMS technique, with ± 0.5 ‰ for *$δ^{18}$*O and ± 2.0 ‰ for *$δ^2$*H values. The development of another analysis system using off-axis integrated cavity output spectroscopy (OA-ICOS) achieved similar precision (Czuppon et al., 2014). The latest analytical systems are able to measure released water volumes in the nano-litre range (50 to 260 nl) with a precision of ± 0.3 ‰ for *$δ^{18}$*O and ± 1.6 ‰ for *$δ^2$*H values using the CRDS technique (Uemura et al., 2016).

The above discussed fluid inclusion extraction lines of Arienzo et al. (2013), Czuppon et al. (2014), and Uemura et al. (2016) are working with dry carrier gas and low water vapour concentrations in the analyser cavity, which may influence the stable isotope measurements by adsorption and cause memory effects. The measured isotopic signal needs to be corrected for, e.g., the isotopic dependence on the water vapour concentration (Uemura et al., 2016) or the memory effect (e.g., van Geldern and Barth, 2012). The memory effect in the analyser cavity is due to limitations during the removal of all gas between two measurements preventing the full desorption of water molecules from the cavity walls. A standard technique to deal with memory effects in liquid water analysis is the repeated injection of the same water sample. The measured signal converges exponentially towards the actual sample signal (e.g., van Geldern and Barth, 2012). However, multiple crushing steps on the same sample are typically not feasible for fluid inclusion measurements of speleothems since the amount of water of these samples is often too low to split it in several aliquots (sub-*μ*l range). The adsorption issue was addressed by Affolter et al. (2014) with an extraction line that is continuously purged with a moist gas providing a water vapour background with constant and known $δ^{18}O$ and $δ^2H$ values. The extraction line with a "wet" $N_2$ gas allows reproducible and precise measurement of released fluid inclusion water. The continuous heating of the system enables the instantaneous evaporation of the water released from inclusions followed by spectroscopic analysis of the resulting mixture of background and sample water vapour. The main advantage of the water-vapour flushing is that this procedure avoids additional corrections of the measured water stable isotopes related to memory effects. The achievable standard deviations of the measurements with this analytical system are smaller than 0.4 ‰ for $δ^{18}O$ and 1.5 ‰ $δ^2H$ values, which is comparable with the traditional IRMS technique (Vonhof et al., 2006; Dublyansky and Spötl, 2009) and CRDS setups working with a dry carrier gas (Arienzo et al., 2013; Czuppon et al., 2014).

One important application of isotope fluid inclusion studies is the reconstruction of paleotemperatures using the oxygen isotope fractionation between water and calcite. The temperature reconstruction was initially based on the measurement of fluid inclusion $\delta^2H$ values from which the fluid inclusion $\delta^{18}O$ values were calculated using the $\delta^2H$-$\delta^{18}O$ relationship of the global meteoric water line (Craig, 1961). Combined with the $\delta^{18}O$ value of the calcite, temperatures were calculated from the oxygen isotope fractionation between the carbonate mineral and water. This indirect approach achieved a reported precision of about ± 2 °C in the early studies of Schwarcz et al. (1976). In the decades prior to 2010, direct temperature calculation from measured fluid inclusion $\delta^{18}O$ values has been rare due to severe challenges in its analysis with standard mass spectrometric approaches. Therefore, mostly the indirect way of first converting measured $\delta^2H$ values into fluid inclusion water $\delta^{18}O$ values was used (e.g., Matthews et al., 2000). More recent studies provided comparable temperature precision based on inclusion water $\delta^{18}O$ values: ± 1.3 °C (van Breukelen et al., 2008), ± 0.9-2.1 °C (Meckler et al., 2015); ± 2.7 °C (Arienzo et al., 2015), and ± 0.6-3.1 °C (Uemura et al., 2016). The uncertainty of the indirect paleotemperature reconstruction from $\delta^2H$ variations in speleothem fluid inclusions is similar: ± 1.5 °C (Zhang et al., 2008) and ± 0.9-2.5 °C (Meckler et al., 2015). More recently, a better precision has been achieved when using the rainfall $\delta^2H$/T relation in mid-latitudes: ± 0.2-0.5 °C (Affolter et al., 2019).

In this study, we systematically assess the measurement method of stable water isotopes in speleothem fluid inclusion analysis using CRDS, including in particular the effect of sample amount (water amount per analysis), water adsorption on freshly crushed calcite surfaces, influence of the isotope values of the water vapour background on the sample signal, as well as the external reproducibility of speleothem fluid inclusion samples (using adjacent aliquots along growth layers). In addition, we present a recent case study allowing to determine paleo-temperature trends in the 1°C range.

## *2. Methods and site description*

### *2.1 Water extraction from fluid inclusions*

At the Institute of Environmental Physics (Heidelberg) water from speleothem fluid inclusions is extracted within a system that is constantly purged by an artificially prepared moist gas, leading to a water vapour background with known $\delta^{18}O$ and $\delta^2H$ values (Fig. 1). In the extraction line this stable water vapour background is generated by mixing water of a known isotopic composition into a dry nitrogen gas flow (300 ml/min). A peristaltic pump (Ismatec -REGLO *Digital*, Wertheim, Germany) continuously supplies small amounts of water (1 μl/min) to the line through a T-injection port with a septum (Fig. 1 A). A constant temperature of 120 °C ensures a complete and immediate evaporation. Instant water evaporation is induced in a fused silica capillary, which slightly touches the heated base of the port. A two-litre mixing cavity placed after the T-injection port generates a stable water vapour background and compensates fluctuations caused by the peristaltic pump cycles. The nitrogen flow is controlled by a mass flow controller (Analyt MTC, model GFC-17, Müllheim, Germany) and creates a constant overpressure of 0.5 bar. The flow rate is 40 ml/min into the CRDS analyser (L2130-i, Picarro, Santa Clara, USA). The surplus gas stream is vented through a purge capillary before the crusher unit. With this setup the water vapour concentration in the CRDS cavity ranges between 6000 and 8000 ppmV, but the cavity can also be adjusted to higher or lower water vapour concentrations if needed. We have chosen a range between 6000 and 8000 ppmV to allow for the detection and analysis of small fluid inclusion water amounts (sufficiently high ratio of water vapour from the sample relative to the background) but also to provide a background water vapour concentration that prevents memory effects.

The sample (speleothem fragment or glass capillary) is inserted in a copper tube (Fig. 1 B) which is connected to the extraction and measurement system. Due to limitation of the copper tube with respect to length and diameter, compact mineral pieces with rectangular dimensions of 6 mm x 6 mm x 10-20 mm are preferred. The copper tube including the sample is purged for at least 30 min until water vapour concentration and isotope values reach a constant signal. The stability of the water vapour background is verified by monitoring the standard deviation of the water vapour concentration. When the standard deviation of the water vapour concentration remained less than 20 ppmV for 30 minutes, the speleothem sample is crushed by compressing the copper tube from the outside with a hydraulic press at 200-300 bar. This compression in the heated system leads to the release and immediate evaporation of inclusion water and a sudden pressure increase. This pressure increase could cause gas flow not only towards the CRDS but also in direction of the purge capillary and may provoke sample gas loss. Therefore, a reflux valve is installed between the 2l mixing cavity and the crushing unit to prevent a backflow and loss of the sample. In general, a very good crushing efficiency has been achieved with an average grain size of 37 μm after the crushing (Weißbach, 2020). A reproducibility test related to the crushing procedure showed similar particle size distributions for five different speleothem samples investigated with laser diffraction (Analysette-22 Micro Tec) after crushing with the hydraulic system.

An injection port is situated next to the copper tube, allowing to mimic a water release from a mineral sample and helps to evaluate and control accuracy and precision of the $δ^{18}O$ and $δ^2H$ values from small (inclusion) water samples. An additional small mixing cavity (400 ml) directly after the crushing unit prolongs the generated sample signal from an usually few seconds lasting peak to a well measurable signal with a duration of several minutes. After the mixing cavity the gas proceeds to the L2130-i isotope and gas concentration analyser (Picarro).

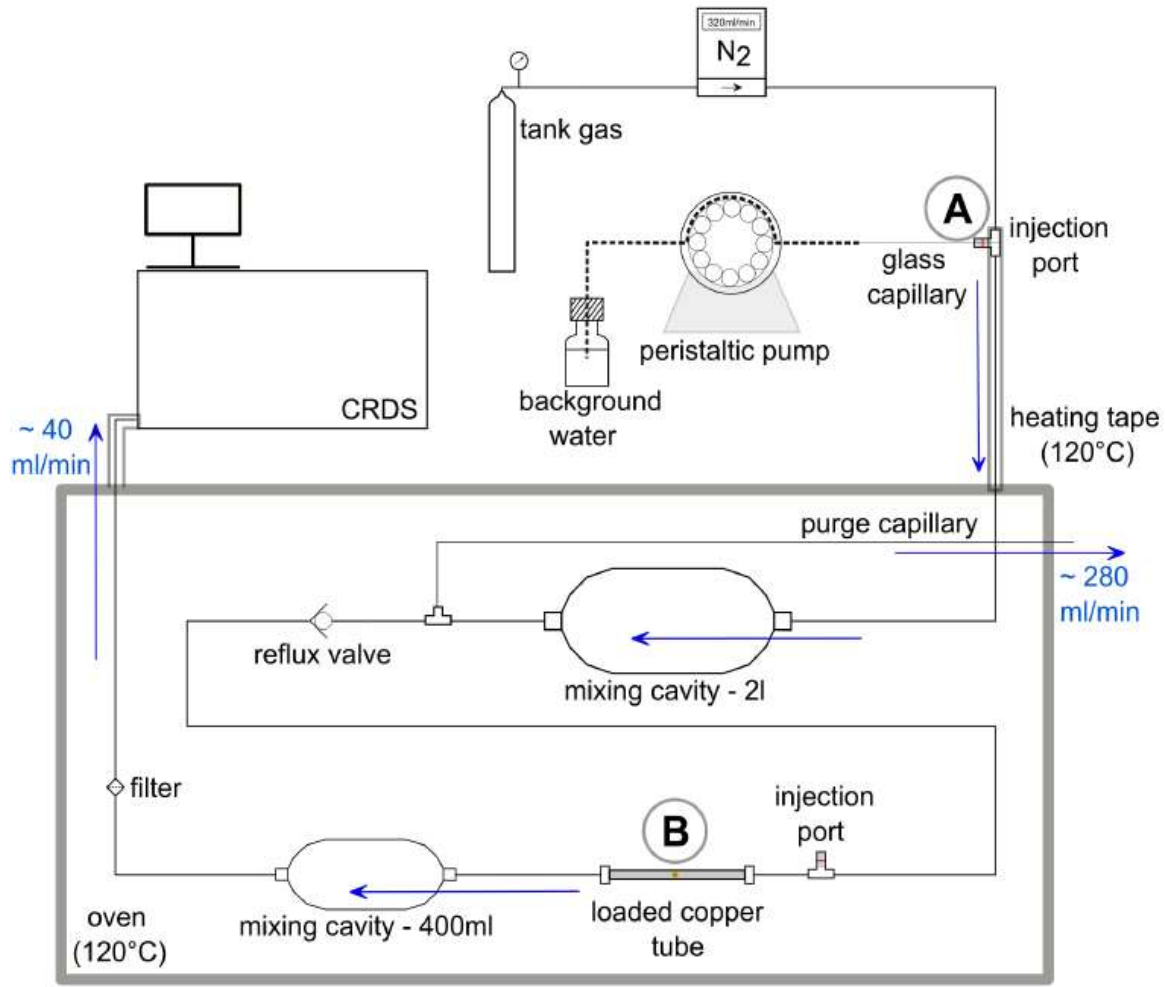

**Figure 1:** Fluid inclusion line for extraction and measurement of stable oxygen and hydrogen isotopes of fluid inclusions in speleothems. Water with a known isotope composition is mixed into a nitrogen gas flow to create a stable water vapour background (position A). The purge capillary reduces the background vapour flow from 280 to 40 ml/min as required for the CRDS analyser (L2130-i, Picarro). The two mixing cavities provide a smoothing of the background vapour signal and a dispersion of the measurement signal. The speleothem sample or the glass capillary is placed in a copper tube and installed at position B inside the heated oven. In order to prevent the backflow of the water vapour from the freshly crushed sample, a reflux valve is installed. The flow directions are indicated as blue arrows. [black/white for figures in print, colour online]

## 2.2 Analysis and data evaluation

The water vapour concentration and the $\delta^{18}O$ and $\delta^2H$ values are determined with the L2130-i isotope and gas concentration analyser of Picarro. The L2130-i analyser is based on wavelength-scanned CRDS in the spectral range from 7183.5 to 7184 $cm^{-1}$ and uses a multi-pass cell that creates a long effective absorption path length of about 12 km (Aemisegger et al., 2012). A stable cavity temperature of 80°C ± 0.002 °C is maintained). The cavity pressure is set to 66.66 hPa (50 Torr). The water isotopologue lines pertaining to $^{18}$O and $^{2}$H are measured simultaneously over a 0.8 s interval. In our setup the L2130i allows isotope measurements in the water vapour concentration range between 1 000 and 50 000 ppmV.

In constant flow mode the actual measured signal is composed of a background signal and a peak signal after crushing and must be integrated over a corresponding time interval (Fig. 2). The evaluation routine follows the approach of Affolter et al. (2014) but was extended by the correction of a potential level change in the water vapour background (Weißbach, 2020). The import of the data and the evaluation was carried out with the assistance of the Python script *IsoFluid (*https://github.com/bhemmer/IsoFluid and http://doi.org/10.5281/zenodo.5911265*)*. *IsoFluid* determines the sample or calibration standard peak start and end based on a slope criterion that compares the start and end slope with the background slope of the water vapour concentration in user-defined time intervals. The sample isotope value ($\delta^{18}O_{sample}$) is calculated by subtraction of the background signal (index back) from the measured signal of the mixed gas signal (index *mix*) and follows the approach of Affolter et al. (2014):

$$\delta^{18}O_{sample} = \frac{\int_{to}^{t1} \delta^{18}O_{sample}(t) * H_2O_{sample}(t) \cdot dt}{\int_{to}^{t1} H_2O_{sample}(t) \cdot dt}$$

$$= \frac{\int_{to}^{t1} \delta^{18}O_{mix}(t) * H_2O_{mix}(t) \cdot dt - \int_{to}^{t1} \delta^{18}O_{back}(t) * H_2O_{back}(t) \cdot dt}{\int_{to}^{t1} H_2O_{mix}(t) \cdot dt - \int_{to}^{t1} H_2O_{back}(t) \cdot dt}$$

(Eq.1)

$H_2O$ refers to the related water amount. The calculation for $\delta^{2}H$ values is correspondent.

All uncertainties are reported at the 1σ level.

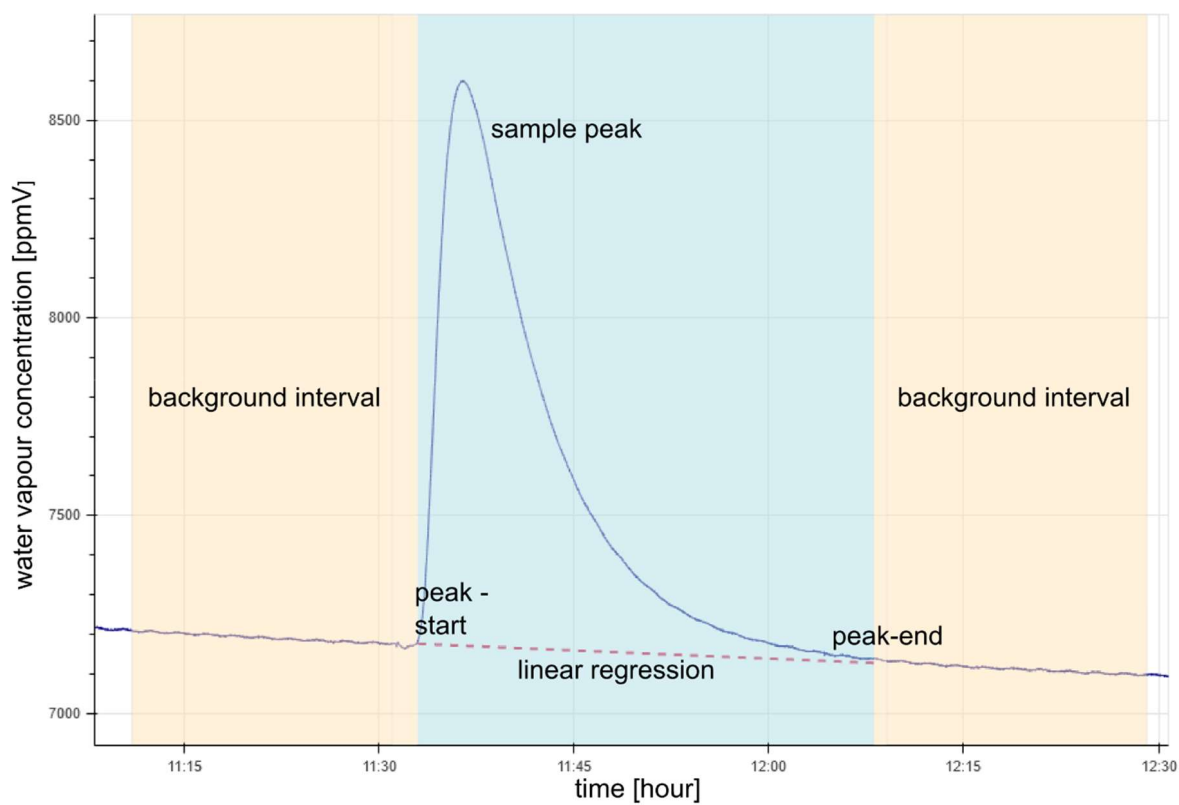

**Figure 2:** Water vapour signal in the CRDS before, during, and after speleothem sample crushing. Background intervals (orange) are used to determine the value of the water background during a sample analysis (blue shaded). [black/white for figures in print, colour online]

## *2.3 Water vapour background calibration*

For the calibration of the oxygen and hydrogen isotope signal of the water vapour background five independently measured in-house reference waters were used (Table 1). Water from Willersinnweiher water (WW) was not used for calibration but for precision and accuracy assessment.

**Table 1**: In - house reference waters for isotopic calibration of the fluid inclusions CRDS system, Isotope values were independently measured at the Institute of Environmental Physics (IUP) at Heidelberg University. Uncertainties are given as 1σ errors.

| Water type | Code | $\delta^{18}O$ values (‰ VSMOW) | $\delta^{2}H$ values (‰ VSMOW) |
|---|---|---|---|
| Artificially evaporated water | AE | 3.8 ± 0.3 | -21.79 ± 1.8 |
| Ocean water | Kona | -0.05 ± 0.08 | 0.5 ± 0.7 |
| Lake Water - Willersinnweiher (S. Germany) | WW | -0.32 ± 0.11 | -18.8 ± 0.4 |
| De-ionized local tap water | VE | -8.57 ± 0.08 | -61.0 ± 0.7 |
| Alpine Water | VCL | -13.04 ± 0.08 | -98.3 ± 0.7 |
| Alpine Water - Colle Gniffeti ice core | CC | -15.13 ± 0.08 | -110.6 ± 0.7 |
| North Greenland water-surface snow | NG | -26.54 ± 0.08 | -212.1 ± 0.7 |

---

The isotope values of these five reference waters were determined independently with a Los Gatos Research LGR1 analyser. These reference waters cover a range of −26.5 up to -0.05 ‰ in $\delta^{18}O$ values and −212.1 up to 0.5 ‰ in $\delta^2H$ values (both VSMOW) (Table 1), which includes the relevant range for speleothem samples. Five different isotope background values were realised by using the corresponding reference water as supply, injected with the peristaltic pump into the system. Once a sufficiently stable water vapour concentration was achieved in the preparation line (standard deviation below 20 ppmV for 30 min) the isotope signal was averaged over 60 minutes, which results in a standard deviation of 0.2 ‰ and 0.7 ‰ for the $\delta^{18}O$ and $\delta^2H$ background values, respectively. Figure 3 shows the CRDS-measured isotope value against the reference value (Table 1). The value of the water vapour background was constantly monitored via repeated measurements of the in-house reference waters, and has remained constant over several years.

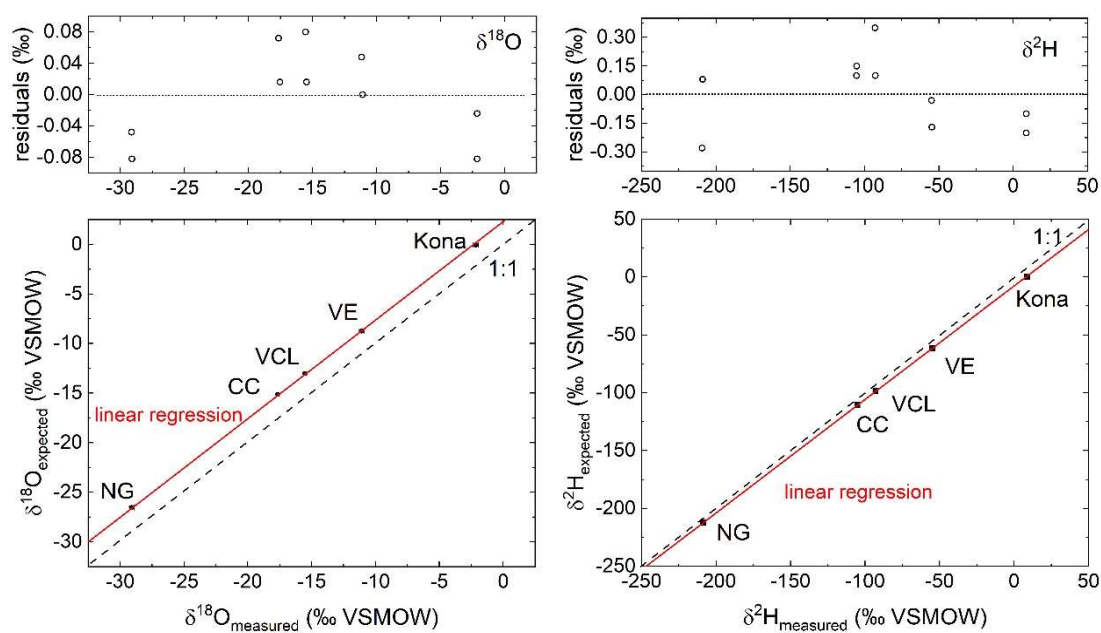

**Figure 3:** The calibration of $\delta^{18}O$ and $\delta^2H$ values of the water vapour background results from a linear regression (red lines). The calibration equation is $y = (0.994 ± 0.007 \cdot x + 2.30 ± 0.14)$ ‰ for $\delta^{18}O$ values ($R^2$=0.999) and $y = (0.980 ± 0.003 \cdot x – 7.77 ± 0.38)$ ‰ for $\delta^2H$ values ($R^2$=0.999). Both calibrations remained constant over several years. Isotope data on the reference waters used for water vapour background calibration are given in Table 1. The residuals from the linear regression indicate that calibrated values are within 0.08 ‰ ($\delta^{18}O$) and 0.35 ‰ ($\delta^2H$) of the expected reference value. Furthermore, the residuals show a random distribution. Uncertainties on the 1σ level in the calibration graphs are smaller than the symbol size. [black/white for figures in print, colour online]

*2.4 **Water amount calibration***

A precise water amount calibration is necessary for determining the exact amount of released water from the crushed speleothem calcite. The released water amount is a major parameter for the calculation of the fluid inclusion isotope value and is determined via water vapour signal integration (see Eq.1). Isotope values could also be calculated with the time-integrated water vapour mixing ratio alone, however, knowledge of the released water amounts is recommended for uncertainty assessment (amount dependence) and for assessment of speleothem growth conditions (fluid inclusion water yield). Typically, volume calibrations are carried out by injecting water in the $\mu$l range with syringes (here: SGE 1BR-7RAX and 5BR-7RAX and Hamilton 70001KH and 75N), however, the variability of the calculated water amount only using syringe injections is significant and can be as high as 10 % (inset in Fig. 4, Weißbach, 2020).

Here we present a water amount calibration method with glass capillaries that follows the approach of Kluge et al. (2008). The glass capillaries (borosilicate, Hirschmann) can be filled with 0.1-5.0 $\mu$l water at varying isotopic composition. They can be closed airtight by melting both ends. The size of the filled capillary can be adjusted to the size of the crushing cell down to a minimum length of approximately 1 cm. The exact volume is determined by scanning the capillary with a high-resolution office scanner and comparison with the pre-marked 1 $\mu$l labels on the capillary. The volume uncertainty of the glass capillary water amount is ± 0.025 $\mu$l and was determined by five repetitions of the manual evaluation of a scan. The accuracy is given by the uncertainty of the pre-marked 1 µl labels (± 0.003 µl). Water-filled capillaries were analysed weekly to monitor the stability of the water amount calibration (Fig. 4). The uncertainty of the water amount determination from the calibration is approximately ± 0.02 µl at a water volume of 1 µl and ± 0.04 µl at 2.5 µl using the 1σ uncertainty of the linear regression. In general, we rarely observed outliers in the water amount calibration when using glass capillaries.

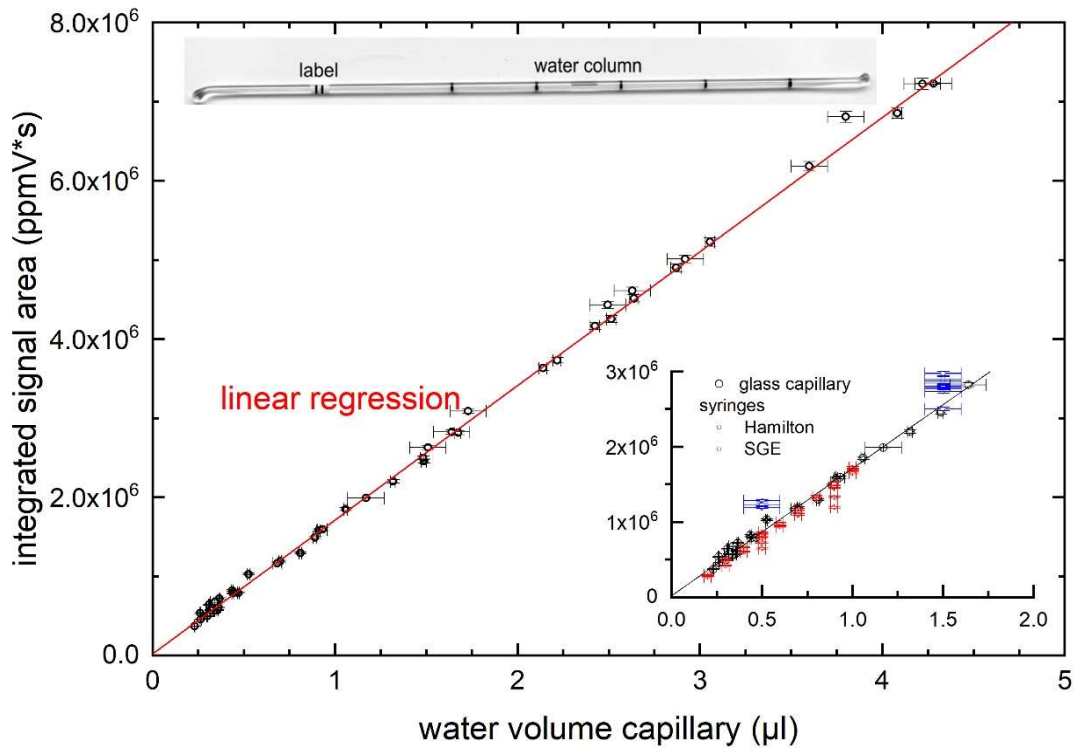

**Figure 4:** The time-integrated measured volume signal in ppmV*s as given by the Picarro analyser is plotted against the water volume of the glass capillaries. The resulting linear regression $y = (5.9 \times 10^{-7} \cdot x - 0.011)$ µl ($R^2=0.999$) is used to determine the released amount of water from speleothem samples. In total 45 capillaries were measured for calibration spanning a water amount range from 0.2 up to 4.3 µl. The upper inset shows a glass capillary filled with about 0.5 µl of water (in the middle of the capillary). The lower inset shows the comparison of water injections with different syringes (blue and red symbols) and the glass capillaries (black circles). The uncertainties are given on the 1σ level. [black/white for figures in print, colour online]

## *2.5 Site and sample description*

### *Hüttenbläserschacht Cave (Germany)*

For direct comparison with rainwater $\delta^{18}O$ values and measured drip water isotope values, we selected a suite of modern and late Holocene samples from the Hüttenbläserschacht Cave, located only a few 100 meters west of the well-monitored Bunker Cave in northwest Germany (e.g., Riechelmann et al., 2011). Both caves are situated in the upper Middle Devonian limestone in Iserlohn (Sauerland). Hüttenbläserschacht Cave hosts pool spar calcites that are expected to provide fluid inclusion isotope values close to drip water as they grow under the water table of the pools. Pool spars from this cave have already been investigated by Kluge et al. (2013) using clumped isotope $\Delta_{47}$ and calcite $\delta^{18}O$ values for calculation of the (drip) water $\delta^{18}O$ value. Calcite was actively precipitating in the pools (e.g., abundant calcite rafts) at the time of pool spar removal. $^{230}$Th-U disequilibrium dating at Heidelberg Academy of Sciences provided radiometric ages of one pool spar and one raft sample of 0.05 ± 0.27 ka BP and 0.36 ± 0.12 ka BP, respectively (Supplemental material S1), corroborating the assumption that the pool spars and rafts are modern age.

*Cloşani Cave (Romania)*

For the second case study we selected a 20[th] century stalagmite (Stam 4) from Cloşani Cave, Romania (e.g., *Constantin and Lauritzen*, 1999). A monitoring program from 2010 to 2012 and 2015 demonstrated a stable cave environment with an air temperature of 11.4 ± 0.5 °C and a relative humidity close to 100 % (*Warken et al.*, 2018). The isotopic composition of the drip water in direct vicinity (1 m) of the former location of Stam 4 showed no seasonal cycle and was constant throughout the monitoring. The mean dripwater $\delta^{18}O$ value was −9.6 ± 0.2‰ and −66.3 ± 1.7‰ for $\delta^2H$ values.

Stalagmite Stam 4 has a total length of 6 cm and an average growth rate of 510 $\mu$m per year, as deduced from counting of layers related to annual cycles in the concentration of various elements (Supplemental material S2). The speleothem grew actively until the removal in spring 2010 C.E. as drip water was feeding the stalagmite. The recent growth of the stalagmite was further constrained by the detection of the 20[th] century radiocarbon bomb spike, which was imprinted by the transport of the atmospheric signal into the speleothem (see Supplemental material S3). Combined layer counting and radiocarbon measurements suggest a growth period from 1910 to 2010 C.E. For the fluid inclusion study, pieces were taken from the peripheral part of Stam 4 with a distance of approximately 1 to 1.5 cm from the actual growth axis (see Supplementary Fig. S2).

### 3. *Results*

#### *3.1 Precision of isotope measurements*

The precision of isotopic measurements (Fig. 5, Table 2) was quantified using the standard deviation of repeated analyses of the reference waters injected via syringes (VE water; Table 1) and independently cross-checked with water-filled glass capillaries using VE and WW reference waters. The injected water amount using syringes) varied between 0.1 and 4.0 $\mu$l.

Using syringe injection method, a clear decrease of the standard deviation of these isotope analyses with increasing water amount becomes apparent (Fig. 5, Table 2). The standard deviation decreases strongest between 0.1 and 1 µL and reaches values between 0.1 and 0.3 ‰ for $\delta^{18}O$ and between 0.2 and 0.7 ‰ for $\delta^2H$ values, for samples larger than 1 µl (Supplementary Table S4). For smaller water amounts, i.e., of 0.5 µl and below, the isotope values of the injections show a significantly larger scatter, leading to standard deviations between 0.4 and 0.5 ‰ for $\delta^{18}O$ and between

1.1 and 1.9 ‰ for $\delta^2H$ values. These uncertainties are based on an exponential fit of the standard deviation against the water volume using repeated measurements at a given water volume (Fig. 5).

For the determination of the precision, reference water sealed in glass capillaries was crushed in the fluid inclusion system. Consistent with the results from the water injections, the precision for isotopic analyses of water released from crushing glass capillaries is between 0.07 and 0.10 ‰ for $\delta^{18}O$ values and between 0.3 and 0.4 ‰ for $\delta^2H$ values, for water amounts above 0.5 µl. Smaller water amounts resulted in a significant increase in the uncertainty and are expressed through a lower precision (Table 2).

**Table 2**: Measurement precision and accuracy in dependence of the water amount. The precision (± 1σ) was determined from repeated injections of isotopically well-characterized water standards using syringes. The accuracy (given at the 1σ level) was assessed by measurement of reference water both from injections and by release from crushing of sealed glass capillaries and comparison with the independently determined isotope values (Table 1). The error estimate for the accuracy assumes a Gaussian distribution and includes the uncertainty of the expected value (VE water: ± 0.08 ‰ for $\delta^{18}O$, ± 0.7 ‰ for $\delta^{2}H$, WW water: ± 0.11 ‰ for $\delta^{18}O$, ± 0.4 ‰ for $\delta^{2}H$). n represents the number of analyses in the investigated water volume range. The water isotope values are given relative to VSMOW.

| Type | Reference water | Water volume (μl) | Precision (1σ) $\delta^{18}O$ value(‰) | $\delta^{2}H$ value (‰) | Accuracy (1σ) $\delta^{18}O$ value (‰) | $\delta^{2}H$ value(‰) | n |
|---|---|---|---|---|---|---|---|
| Water injection | VE | | | | | | |
| | | 0.1 | 0.54 | 1.8 | | | 6 |
| | | 0.2 | 0.34 | 1.6 | | | 6 |
| | | 0.3 | 0.49 | 1.6 | | | 6 |
| | | 0.4 | 0.53 | 1.1 | | | 6 |
| | | 0.5 | 0.58 | 1.4 | | | 6 |
| | | 1 | 0.17 | 0.4 | | | 16 |
| | | 2 | 0.21 | 0.4 | | | 11 |
| | | 3 | 0.16 | 0.4 | | | 11 |
| | | 4 | 0.10 | 0.1 | | | 9 |
| | | Mean all | 0.35 | 1.0 | 0.09 ± 0.19 | 0.3 ± 0.7 | 77 |
| | | Mean < 1μl | 0.50 | 1.5 | 0.12 ± 0.22 | 0.3 ± 0.7 | 30 |
| | | Mean ≥ 1μl | 0.16 | 0.3 | 0.08 ± 0.12 | 0.3 ± 0.7 | 47 |
| Glass capillary | | | | | | | |
| | WW | 0.3-4.3 | 0.42 | 0.4 | 0.10 ± 0.44 | 0.1 ± 0.6 | 16 |
| | WW | > 0.5 | 0.07 | 0.3 | -0.02 ± 0.13 | 0.1 ± 0.5 | 14 |
| | VE | 0.3-4.1 | 0.30 | 1.3 | 0.14 ± 0.31 | 0.6 ± 1.5 | 17 |
| | VE | > 0.5 | 0.10 | 0.4 | -0.03 ± 0.13 | 0.1 ± 0.8 | 9 |

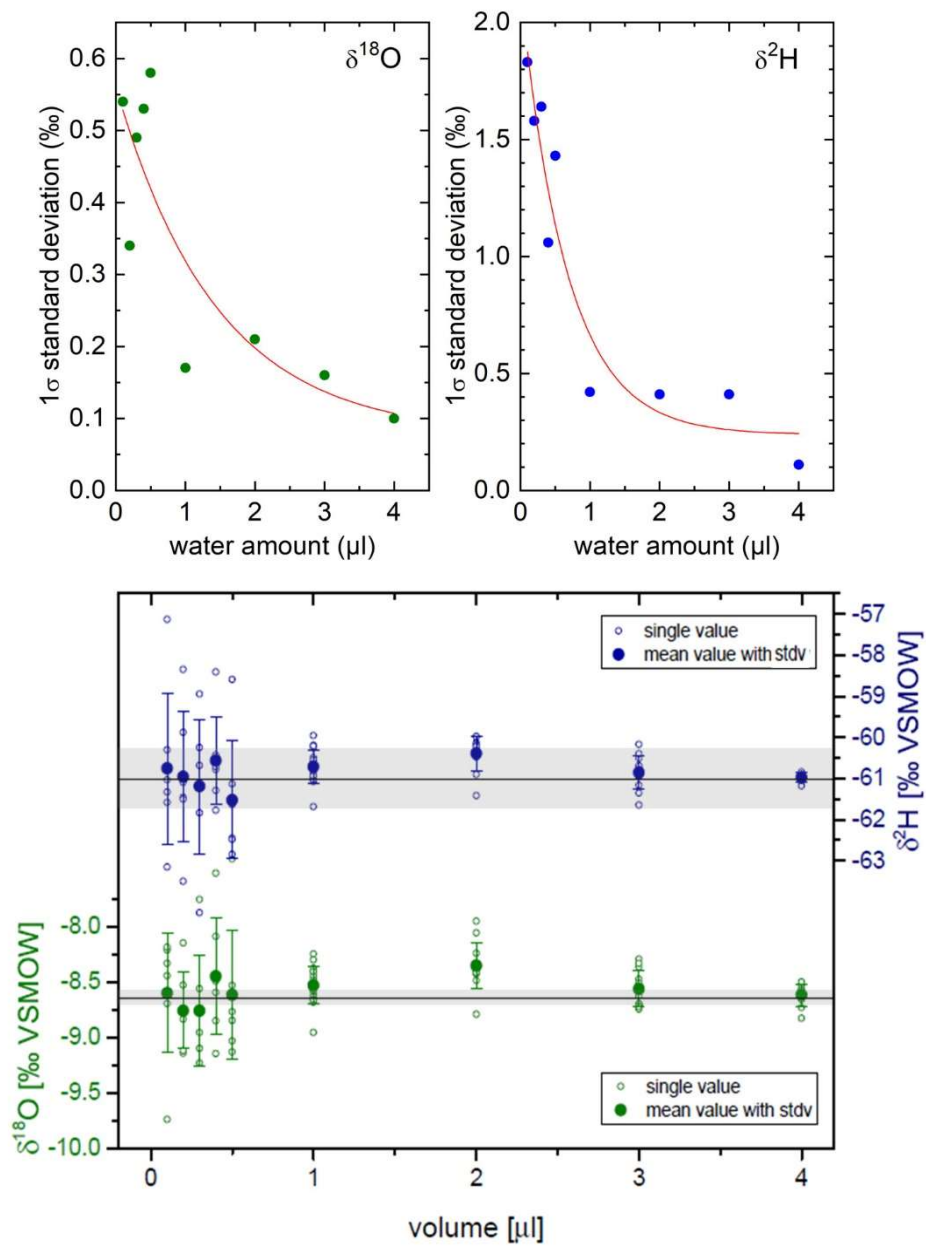

Figure 5: Upper panels: precision (1σ) of the isotope measurements for varying amounts using the water injection method. The red lines represent the least-square exponential fits to the data. The standard deviation decreases with increasing water amount for both $\delta^{18}O$ and $\delta^2H$ values. . Lower panel: accuracy determination for varying water amounts based on individual injections (open circles). The related mean values with their 1σ standard deviation are shown as filled dots with error bars. The black horizontal lines represent the reference value for *VE* - tap water ($\delta^{18}O$=- 8.57 ‰, $\delta^2H$ =- 61.0 ‰) with its uncertainty band (grey shading, ± 0.08‰ for $\delta^{18}O$, ± 0.7 ‰ for $\delta^2H$ values). [black/white for figures in print, colour online]

### 3.2 *Accuracy of isotope analysis of micro-litre water amounts*

The accuracy of the water $\delta^{18}O$ and $\delta^{2}H$ values was assessed for reference waters (Table 1) by the injection with syringes and by crushing glass capillaries. The injected water amounts covered the typical range of water extracted from inclusions (Fig. 5). The glass capillaries were filled with reference water with similar water amounts between 0.3 and 4.3 µl and were crushed in the copper tube with the same hydraulic press as the stalagmite samples.

Considering the water isotope mean values of all measurements performed with the glass capillaries, the $\delta^{18}O$ value deviated from the expected reference values (Table 1) by 0.10 ± 0.44 ‰ for WW water (n=16) and by 0.14 ± 0.31 ‰ for VE water (n=17) (Table 2). For $\delta^{2}H$ values the deviation from the reference value was 0.1 ± 0.6 ‰ for WW water (n=16) and 0.6 ± 1.50 ‰ for VE water (n=17). Considering only those measurements with water amounts above 0.5 µl reduces the uncertainty. For this selection, the $\delta^{18}O$ value of both reference waters deviates on average from the expected reference value by -0.02 ± 0.13 ‰ for WW water (n=14) and -0.03 ± 0.13 ‰ for VE water (n=9). For $\delta^{2}H$ values the deviation from the reference value was 0.1 ± 0.5 ‰ for WW water (n=14) and 0.1 ± 0.8 ‰ for VE water (n=9).

The accuracy as determined by crushing of water-filled glass capillaries is confirmed by the injection-based data (Table 2). Overall, the $\delta^{18}O$ value of the injected VE water deviated from the expected value on average by 0.09 ± 0.19 ‰ (n=77), that of the $\delta^{2}H$ value by 0.3 ± 0.7 ‰ (n=77).

### 3.3 *Adsorption and/or desorption on the calcite surface*

Adsorption on a calcite surface and, in particular, on freshly crushed carbonate with a large surface to volume ratio provides the possibility to alter the isotope values of the fluid inclusion water (Dennis et al., 2001). Therefore, an artificial fluid inclusion system (speleothem analogue) as described by Dennis et al. (2001) has been prepared to quantify the influence of adsorption on the measured isotopic signal in our setup. We measured water vapour released from a water-filled glass capillary in direct contact with inclusion-free Iceland spar carbonate. The compact Iceland spar pieces (0.45-0.81 g) as well as the released water of the capillaries (1.4-3.7 µl water) represent a speleothem sample with a water content of 2.2 up to 7.8 $\mu$l per g calcite. In total, we prepared and analysed five artificial fluid inclusion - calcite systems in the range between 5.2 and 7.8 µl/g and three at about 2.2 µl/g and compared them to water-filled glass capillaries without additional calcite. The crushing of the compact Iceland spar pieces provided fresh and fine-grained calcite for interaction and adsorption testing. The measurements suggest that the adsorption of water molecules on the calcite surfaces does not affect the measured isotopic signal in the investigated water/calcite ratio range (Fig. 6). Both measured oxygen and hydrogen isotope values accurately match the expected value. With a standard deviation of ± 0.05 ‰ for $\delta^{18}O$ and ± 0.22 ‰ for $\delta^{2}H$ values (high water/calcite ratio, n=5) and ± 0.15 ‰ for $\delta^{18}O$ and ± 0.31 ‰ for $\delta^{2}H$ values (low water/calcite ratio, n=3) in both adsorption tests, a good reproducibility of the individual measurements was achieved. We observed that after crushing of Iceland spar (0.25 g) 0.023 µl water was adsorbed on the crushed calcite from the moist carrier gas (Supplementary Fig. S3), which corresponds to a ratio of approximately 0.1 µl water per g calcite. Thus, for low water contents of < 0.1 µl per g calcite an influence of adsorption on the released water amount and the isotopic values probably cannot be excluded. Therefore, we rejected all fluid inclusion samples with water amounts below 0.1 µl based on this observation (see Weißbach, 2020).

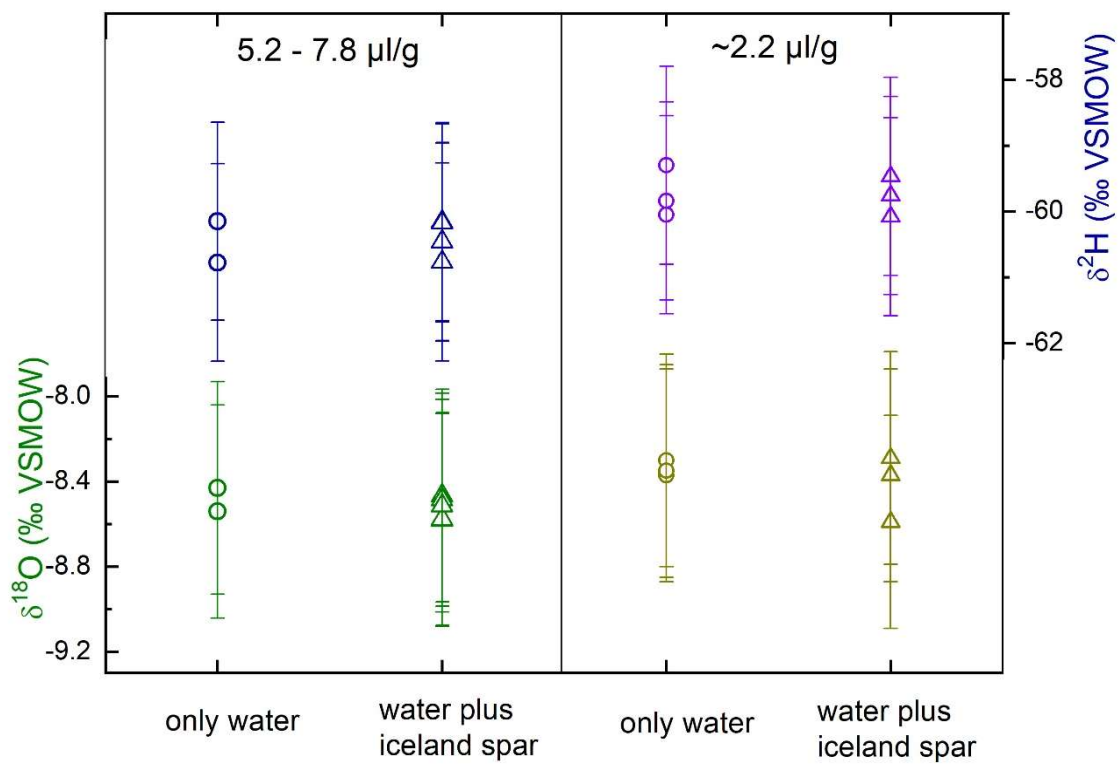

**Figure 6:** Isotopic values measured for the artificial inclusion calcite system, for which compact Iceland spar pieces were crushed together with *VE* water - filled glass capillaries (triangles). Open circles indicate water - filled glass capillaries (*VE*) without calcite addition, measured for comparison. An isotopic fractionation due to adsorption of water molecules on the calcite surface is not detectable for the investigated water content range of 5.2-7.8 µl water/g calcite (left side) and for 2.22±0.08 µl water/g calcite (right). Marginal differences in the isotope values between left and right panel are within the expected variations in the reference water isotope values due to a several year time lag between both experimental series. The uncertainties are displayed on the 1σ level. [black/white for figures in print, colour online]

### *3.4 Isotopic effect of the water vapour background*

The potential influence of the isotope ratio of the water vapour background on that of the measured sample could be relevant for speleothem samples whose isotopic composition strongly differs from that of the water vapour background. For testing this potential effect, we injected our VE water standard on four different water vapour backgrounds with different isotopic composition (Fig. 7). We used VE-water as injection fluid, because its isotopic composition is comparable to the majority of fluid inclusion of speleothems from mid-latitudes. For each water vapour background 3.0 $\mu$l of VE water were injected five times. For the background waters with the two most extreme isotope compositions we additionally injected 1.0 µl of VE (n=6) to assess the robustness also for smaller water amounts. .

If *VE* water is injected on *VE* background water vapour, the average isotope value corresponds to the expected value within uncertainty. A deviation from the expected isotope value is notable for injections on a different water vapour background. For example, VE injections on a negative water vapour background (NG, $\delta^{18}O$ = -26.54 ± 0.08 ‰, $\delta^2H$ = -212.1 ± 0.7 ‰, Table 1) yield a deviation of +0.40 ‰ for $\delta^{18}O$ and +2.9 ‰ for $\delta^2H$ values from the reference values. VE injections on a background, which is based on lake water (WW) with higher isotope values compared to VE water, yield deviations of -0.15 ‰ for $\delta^{18}O$ and -0.3 ‰ for $\delta^2H$ values. Tests with injected water amounts of 1 µl corroborate the observed trend (Fig.7). The standard deviation of repeated water injections is independent from the isotopic composition of the water vapour background. The effect of the isotopic difference between the samples and the background water vapour exceeds the measurement uncertainty only for differences larger than 10 ‰ ($\delta^{18}O$). ). This experiment highlights that it is not necessary to correct samples when using background water with an isotope composition close to the paleoclimate samples. For speleothem measurements we used VE water as background water.

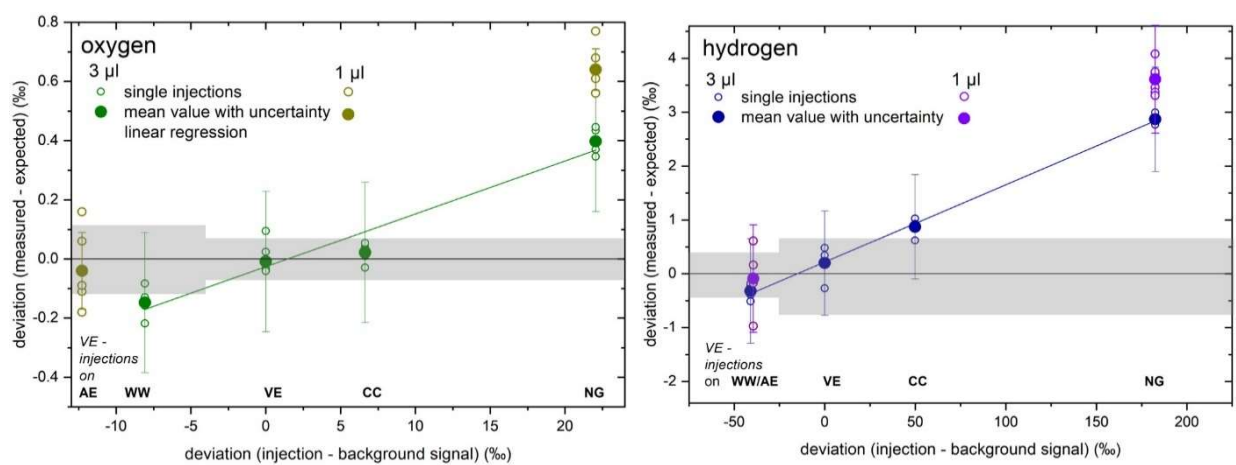

**Figure 7**: Deviation of the measured injection isotope value relative to the expected value. The deviation is related to the difference between the isotope signal of the injection and that of the water vapour background. Single injections are shown as open circles and the mean values as filled circles s. The green andblue lines indicate the linear regression with all individual 3 µl measurements. An increasing deviation between the measured and expected isotopic signal is observed for an increasing difference between injection isotope value and that of the water vapour background. The uncertainty of the expected value (black horizontal line) is shown as grey envelope. Measurement uncertainties are given on the 1σ level. [black/white for figures in print, colour online]

### *3.5 Case applications*

### *Case example 1: Modern – late Holocene sinter samples*

The two fluid inclusion replicates of each modern or late Holocene sample from Hüttenbläserschacht Cave reproduce very well and are within uncertainty of each other (Table 3). Related standard deviations of the mean (0.2 and 1.6 ‰ for $\delta^{18}O$ and $\delta^{2}H$ values, respectively) are comparable ($\delta^{18}O$ values) or slightly larger ($\delta^{2}H$ values) than the measurement precision in this water amount range (0.5-3.0 µl, Fig. 5). The mean fluid inclusion $\delta^{18}O$ value of -7.6 ± 0.2 ‰ is identical to the calculated drip-water value of Kluge et al. (2013) of -7.6 ± 0.3 ‰, independently confirming the former finding. Drip water in Hüttenbläserschacht Cave was not monitored but should be close to the neighbouring Bunker Cave and shares the same karst aquifer with comparable residence times of a few years (e.g., Kluge et al., 2010). Fluid inclusion isotope values are close to the mean drip water values from Bunker Cave of -7.9 ± 0.2 ‰ for $\delta^{18}O$ and -53.3 ± 1.6 ‰ for $\delta^{2}H$ values (Riechelmann et al., 2017).

**Table 3**: Measurement of fluid inclusions in three $CaCO_3$ spar samples from Hüttenbläserschacht Cave (Germany). Each sample was split in two to allow for a replication test. 'Avg.' refers to the average of the two analyses. For comparison also the calculated pool water $\delta^{18}O$ value of Kluge et al. (2013) is shown that uses an independent temperature estimate and clumped isotope $\Delta_{47}$ for correction of kinetic isotope effects. The Bunker Cave drip water is taken from Riechelmann et al. (2017). Uncertainties are given on the 1σ level.

|  | ID | Sample weight (g) | Water amount (µl) | Water content (µl/g) | $\delta^{18}O$ value (‰ VSMOW) | $\delta^{2}H$ value (‰ VSMOW) |
|---|---|---|---|---|---|---|
| Pond A | A-1 | 0.52 | 0.24 | 0.46 | -7.5 ± 0.5 | -53.2 ± 1.5 |
|  | A-2 | 0.52 | 0.31 | 0.60 | -7.9 ± 0.5 | -51.9 ± 1.5 |
|  | Avg. | - | - | - | -7.7 | -52.5 |
| Pond B | B-1 | 0.57 | 0.43 | 0.76 | -7.6 ± 0.5 | -49.7 ± 1.5 |
|  | B-2 | 0.65 | 0.42 | 0.64 | -7.8 ± 0.5 | -48.6 ± 1.5 |
|  | Avg. | - | - | - | -7.7 | -49.1 |
| Pond C (little pond) | C-1 | 0.59 | 1.78 | 3.02 | -7.3 ± 0.3 | -51.1 ± 1.0 |
|  | C-2 | 0.45 | 0.48 | 1.08 | -7.7 ± 0.5 | -51.0 ± 1.5 |
|  | Avg. |  |  |  | -7.5 | -51.0 |
| Average all |  |  |  |  | -7.6 ± 0.2 | -50.9 ± 1.6 |

| | | | |
|---|---|---|---|
| Reconstructed after Kluge et al. (2013) | | -7.6 ± 0.3 | |
| Drip water Bunker Cave | | | |
| range 2006-2013 | | -8.5 to -7.0 | -48 to -58 |
| mean 2006-2013 | | -7.9 ± 0.2 | -53.3 ± 1.6 |

### *Case example 2: Speleothem sample from the 20<sup>th</sup> century – Stam 4 from Cloşani Cave*

*Comparison with current drip water and reproducibility assessment*

We sampled calcite pieces at the outer surface of the stalagmite for comparison with current drip water. It can be assumed that recent calcite precipitated there and accordingly, recent drip water is enclosed in the fluid inclusions. The water yields during crushing were between 0.49 and 1.38 µl/g with a mean of 0.93 ± 0.28 µl/g (one sample was excluded due to a low water amount of 0.18 µl) (Supplementary Table S3). The mean value of 13 fluid inclusion measurements of samples from the outer stalagmite layer is $\delta^{18}O$ = −9.5 ± 0.5 ‰ and $\delta^{2}H$ = −64.6 ± 1.2 ‰ (Supplementary Table S3). These values agree within uncertainty with the mean of the related drip site CL3 of $\delta^{18}O$ = −9.6 ± 0.2 ‰ and $\delta^{2}H$ = −66.3 ± 1.7 ‰ (Fig. 8). The 13 individual measurements reproduce with a standard deviation of 0.5 ‰ and 1.2 ‰ for $\delta^{18}O$ and $\delta^{2}H$ values, respectively, which is slightly higher than the analytical uncertainty based on the standard deviation of repeated syringe injection for water amounts between 0.5-and 1.7 µl (0.2-0.4 ‰ for $\delta^{18}O$ values, 0.4-1.1 ‰ for $\delta^{2}H$ values, Fig. 5). The standard deviation for the 13 individual speleothem analyses is also comparable to that of other CRDS systems and similar water amount ranges, such as of Arienzo et al. (2013) with ± 0.5/2.0 ‰ for $\delta^{18}O/\delta^{2}H$ values and Affolter et al. (2014) with ± 0.5/1.5 ‰ for $\delta^{18}O/\delta^{2}H$ values . The precision of the Stam4 sample analysis also compares well with traditional IRMS measurement techniques which achieve a precision of ± 0.5 ‰ for $\delta^{18}O$ and ± 2.0 ‰ for $\delta^{2}H$ values for water amounts > 0.2 µL (Dublyansky and Spötl, 2009).

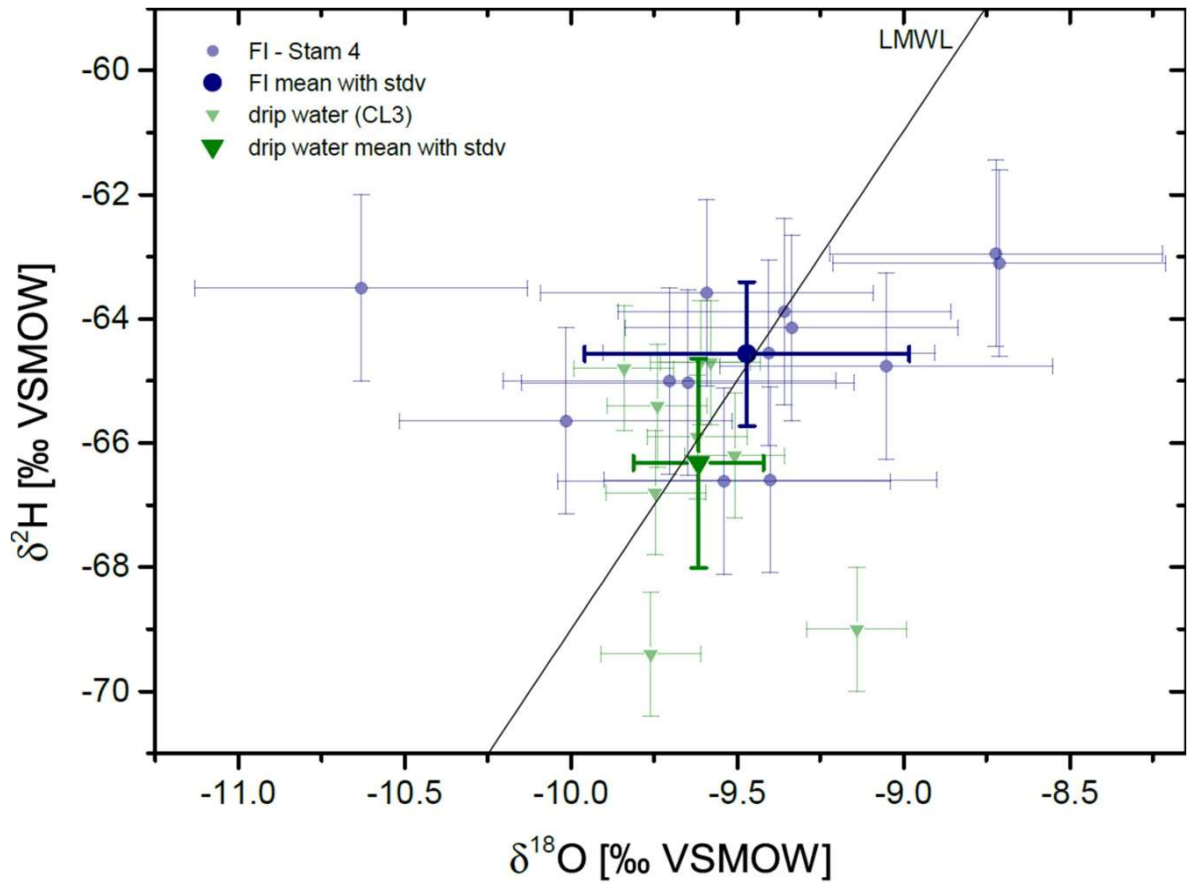

**Figure 8:** Fluid inclusion water isotope ratios of samples from the outer stalagmite surface (light blue dots), with corresponding mean value (dark blue). Drip water data from the same cave chamber where Stam 4 was removed (light green triangles; drip site CL3) and its mean value (dark green) agree with the fluid inclusion results. Both drip water and fluid inclusion data match the local meteoric water line (LMWL) of Cluj-Napoca of $\delta^2H = 8.03 \cdot \delta^{18}O + 11.29$ ‰ (*Cozma et al.*, 2017). The uncertainties are given on the 1σ level. [black/white for figures in print, colour online]

*Fluid inclusion analysis of samples along the growth axis*

We used the stalagmite pieces closest to the growth axis of Stam 4 for paleo-drip water and -temperature reconstruction (Fig. S2). Where possible, the reproducibility of the individual measurements was tested with a second set of fluid inclusion samples, extracted adjacent to the first set of samples (Table 4). The second set had a larger distance from the growth axis than the first set. The samples corresponding to the same growth period are grouped in levels, indicated by letters A-K (Fig. 9). For sample level D, only the second sample is used because the first sample is close to the applied water amount limit and contains only 0.18 μl. On average, the $\delta^2H$ values of sample and replicate are largely consistent (mean deviation: 0.1 ± 0.8 ‰). The same is observed for the inclusion water $\delta^{18}O$ value (mean deviation: 0.31 ± 0.51 ‰). In addition the water content of the different levels appears characteristic. For level D and E with 5 replicates each, the water content varies only 0.1 μl/g (excluding one sample each with low total water amount). For the other levels, a higher scatter has been observed, potentially due to a general heterogeneity of the speleothem inclusion distribution (e.g., Muñoz-García et al., 2012). Generally, the water content was between 0.45 and 1.66 µl/g, suggesting minimal or negligible influence of adsorption on the freshly crushed surface (Table 4). Fluid inclusion $\delta^{18}O$ values vary between -10.4 ‰ and -8.0 ‰ and, with one exception (level C), follow a temporal trend towards higher values towards more recent times (Fig. 9, Table 4).

**Table 4**: Measurement results of fluid inclusion samples of stalagmite Stam-4 from Cloşani Cave (Romania). Several samples were cut from individual layers that reflect contemporaneously grown carbonate and allow for replication tests. The fractionation factor $^{18}\alpha(CaCO_3\text{-}H_2O)$ between water and $CaCO_3$ was calculated based on the difference between the calcite $\delta^{18}O$ values (averaged over the edge length of the fluid inclusion sample of typically 5 mm) and the fluid inclusion water $\delta^{18}O$ values. The temperature $T_{18O,cc}$ was determined using the $^{18}\alpha(CaCO_3\text{-}H_2O)$ - T relationship proposed by Kim and O'Neil (1997). $T_H$ is related to the relative temperature change calculated using the $\delta^2H$-temperature relationship in rainfall (4.72‰/°C) and was referenced to top level K and the current cave temperature. $T_{18O,Fi}$ refers to the temperature difference relative to sample level K with the modern cave temperature as reference and was calculated using the $\delta^{18}O$-T relationship in rainfall (0.59‰/°C). Samples in grey are not included in the interpretation and discussion as the water amount was 0.19 µl or below. Samples closest to the growth axis ('1' closest, higher numbers are further away) were used for temperature assessment based on the classical carbonate thermometer. Samples A1 and A2 were the oldest samples and were excluded from the discussion as they belong to the stalagmite base with unclear chronology. The age corresponds to the mean age of each sample level. Dft: distance from top.

| ID | Dft (mm) | Age (year AD) | Sample weight (g) | Water amount (µl) | Water content (µl/g) | $\delta^2H$ value (‰ VSMOW) | $\delta^{18}O$ value (‰ VSMOW) | $^{18}\alpha$ (CaCO$_3$-H$_2$O) (‰) | $T_{18O,cc}$ (°C) | $T_{18O,Fi}$ (°C) | $T_H$ (°C) |
|---|---|---|---|---|---|---|---|---|---|---|---|
| A1 | | unknown | 0.58 | 0.30 | 0.52 | -65.4 ± 1.5 | -9.5 ± 0.5 | | | | |
| A2 | | | 0.49 | 0.29 | 0.59 | -59.7 ± 1.5 | -9.8 ± 0.5 | | | | |
| B1 | 48.2 | 1928 | 0.42 | 0.40 | 0.95 | -64.1 ± 1.5 | -9.6 ± 0.5 | 31.8 ± 0.5 | 7.7 ± 2.2 | 9.4 ± 0.2 | 10.4 ± 0.5 |
| B2 | | | 0.49 | 0.37 | 0.75 | -64.7 ± 1.5 | -9.6 ± 0.5 | | | | |
| B3 | | | 0.53 | 0.40 | 0.76 | -64.8 ± 1.5 | -8.9 ± 0.5 | | | | |
| B4 | | | 0.52 | 0.29 | 0.56 | -66.3 ± 1.5 | -9.1 ± 0.5 | | | | |
| B5 | | | 0.42 | 0.19 | 0.45 | -68.5 ± 1.5 | -9.4 ± 0.5 | | | | |
| C1 | 43.9 | 1937 | 0.49 | 0.81 | 1.66 | -57.6 ± 1.0 | -8.0 ± 0.3 | 30.7 ± 0.5 | 12.7 ± 2.3 | 12.1 ± 0.1 | 11.8 ± 0.4 |
| C2 | | | 0.42 | 0.51 | 1.21 | -59.6 ± 1.0 | -8.5 ± 0.3 | | | | |
| C3 | | | 0.54 | 0.44 | 0.81 | -60.4 ± 1.5 | -9.0 ± 0.5 | | | | |
| D1 | | | 0.32 | 0.18 | 0.57 | -63.8 ± 1.5 | -8.5 ± 0.5 | | | | |
| D2 | 39.0 | 1944 | 0.51 | 0.42 | 0.83 | -63.8 ± 1.5 | -10.0 ± 0.5 | 32.6 ± 0.5 | 4.3 ± 2.1 | 8.7 ± 0.2 | 10.5 ± 0.5 |
| D3 | | | 0.56 | 0.50 | 0.89 | -63.7 ± 1.5 | -10.4 ± 0.5 | | | | |

| | | | | | | | | | | | | | |
|---|---|---|---|---|---|---|---|---|---|---|---|---|---|
| D4 | | | 0.55 | 0.54 | 0.98 | -64.0 ± 1.5 | -10.1 ± 0.5 | | | | | | |
| D5 | | | 0.40 | 0.35 | 0.88 | -61.7 ± 1.5 | -9.0 ± 0.5 | | | | | | |
| E1 | 34.5 | 1952 | 0.49 | 0.38 | 0.78 | -62.9 ± 1.5 | -10.3 ± 0.5 | 32.7 ± 0.5 | 3.5 ± 2.1 | 8.2 ± 0.2 | 10.7 ± 0.5 |
| E2 | | | 0.58 | 0.46 | 0.78 | -62.5 ± 1.5 | -9.4 ± 0.5 | | | | |
| E3 | | | 0.54 | 0.44 | 0.82 | -62.8 ± 1.5 | -10.4 ± 0.5 | | | | |
| E4 | | | 0.53 | 0.32 | 0.60 | -63.1 ± 1.5 | -10.0 ± 0.5 | | | | |
| E5 | | | 0.25 | 0.14 | 0.55 | -59.4 ± 1.5 | -9.0 ± 0.5 | | | | |
| F1 | 30.1 | 1960 | 0.47 | 0.35 | 0.75 | -63.5 ± 1.5 | -9.6 ± 0.5 | 31.9 ± 0.5 | 7.0 ± 2.2 | 9.4 ± 0.2 | 10.5 ± 0.5 |
| F2 | | | 0.58 | 0.82 | 1.4 | -63.2 ± 1.0 | -9.4 ± 0.3 | | | | |
| F3 | | | 0.56 | 0.87 | 1.56 | -59.8 ± 1.0 | -8.3 ± 0.3 | | | | |
| G1 | 26.2 | 1968 | 0.46 | 0.40 | 0.87 | -62.0 ± 1.5 | -9.0 ± 0.5 | 31.3 ± 0.5 | 9.7 ± 2.2 | 10.4 ± 0.2 | 10.9 ± 0.5 |
| G2 | | | 0.50 | 0.76 | 1.53 | -61.8 ± 1.0 | -8.9 ± 0.3 | | | | |
| H1 | 22.1 | 1977 | 0.40 | 0.46 | 1.16 | -61.6 ± 1.0 | -9.1 ± 0.3 | 31.9 ± 0.5 | 7.2 ± 2.2 | 10.2 ± 0.1 | 10.9 ± 0.4 |
| H2 | | | 0.35 | 0.46 | 1.30 | -61.0 ± 1.0 | -8.8 ± 0.3 | | | | |
| I | 17.8 | 1990 | 0.39 | 0.28 | 0.72 | -60.3 ± 1.5 | -9.3 ± 0.5 | 32.0 ± 0.5 | 6.8 ± 2.2 | 9.9 ± 0.2 | 11.2 ± 0.5 |
| J | 10.9 | 2004 | 0.50 | 0.43 | 0.86 | -59.2 ± 1.5 | -8.7 ± 0.5 | 31.3 ± 0.5 | 9.9 ± 2.2 | 10.9 ± 0.2 | 11.4 ± 0.5 |
| K | 3.9 | 2008 | 0.46 | 0.43 | 0.94 | -59.4 ± 1.5 | -8.4 ± 0.5 | 30.7 ± 0.5 | 12.3 ± 2.3 | 11.4 ± 0.2 | 11.4 ± 0.5 |

## *4. Discussion*

### *4.1 Constraints for precise and accurate fluid inclusion isotope data*

The presented setup allows for a good reproducibility with respect to isotope measurements of pure water samples in the µl range, either injected via a syringe or by crushing of water-filled glass capillaries in the copper tube (similar to speleothems samples; section 3.3). The achievable precision is 0.4-0.5 ‰ for $\delta^{18}O$ and 1.1-1.9 ‰ for $\delta^2H$ analyses at extracted water amounts between 0.1 µl and 0.5 µl and decreases with increasing water amount to ± 0.1-0.3 ‰ for $\delta^{18}O$ and ± 0.2-0.7 ‰ for $\delta^2H$ measurements at extracted water amounts >1 µl (Fig. 5). The improved precision with increasing water amount is consistent with the observations of Dassié et al. (2018) who reported similar precision of 0.2-0.3 ‰ for $\delta^{18}O$ and 0.6-2.6 ‰ for $\delta^2H$ values for 0.2-1 µl as well as a strong increase of the uncertainty at water amounts lower than 0.1 µL. Replicate analyses of calcite samples from the outermost surface of a Romanian stalagmite corroborate the precision as determined by crushing of water-filled glass capillaries and water injections.

Adsorption of water on freshly crushed surfaces appears negligible for water contents of about 1 µl water per g calcite or above Dennis et al. (2001) similarly observed a decreasing adsorption influence at increasing $H_2O/CaCO_3$ ratios at room temperature. However, an adsorption effect could be relevant if the water content in the crushed samples approaches 0.1 µl/g or is below this value. We therefore recommend to use the water content as one parameter to check the robustness of the analysis and to carefully assess or conservatively reject samples with water contents below 0.1 µl/g.

We observed a small dependence of the measured isotope value on the water vapour background (Fig. 7).. After injection of a certain water amount, the $\delta^{18}O$ value of the (hypothetically) well mixed water vapour consisting of background and injection water is an amount-weighted mixture of both $\delta^{18}O$ values. For background water with relatively depleted values such as North Greenland Water (NG, -$\delta^{18}O$ =-26.5 ‰ and $\delta^2H$ = -212.1 ‰) this would mean that the $\delta^{18}O$ value of the VE water with $\delta^{18}O$ = -8.57 ‰ and $\delta^2H$ = -61.0 ‰ is higher than the mixed water. For example, if the background to injection volume is 1.8:3.0, the isotopic composition of the mixture is expected to be $\delta^{18}O$ = -15.3 ‰ and $\delta^2H$ = -117.7 ‰.Given the short residence time of the water vapour in the mixing cavity before the measurement in the CRDS, a full isotopic mixing is not reached. The kinetically slower molecules containing an $^{18}O$ atom remain preferentially in the gas stream compared to the faster molecules containing only $^{16}O$ atoms that preferentially take part in the mixing with the background water. Thus, for this case example it is expected that the injection water isotopes are slightly higher relative to the background and the mixed signal. Conversely, for a positive background as the WW water, the isotope value of the VE injection is more negative relative to the isotope value of the hypothetical fully mixed gas stream. Due to the kinetic behaviour of $^{18}O$, the injection stays more negative relative to the expected value for this background. The adsorption effects and the influence of kinetic isotope exchange are similar for the 1 µl and 3 µl injections (Fig. 7).

For water amounts in the µl range this dependence on the vapour background isotope value is relevant if the isotopic composition of the fluid inclusions is significantly different from the background (> 10 ‰ for $\delta^{18}O$ and > 50 ‰ for $\delta^2H$). Otherwise, the potential effect of the isotopic difference to the background water vapour is within the analytical uncertainty of water samples between 0.1 and 1.0 µl. The maximum expected deviations are < 0.25 ‰ for $\delta^{18}O$ and < 1.0 ‰ for $\delta^2H$ values, if the sample is within the 10 ‰ range of the water vapour background for $\delta^{18}O$ and 50 ‰ for $\delta^2H$ values. For water amounts larger than 1 µl the acceptable deviation between sample and background water isotope values reduces in relation to the higher measurement precision at higher water amounts (Fig. 5).

### *4.2 Paleotemperature calculation from Stam 4 using fluid inclusion isotopes*

For calculation of the $CaCO_3$-$H_2O$ isotope fractionation, we averaged the calcite $\delta^{18}O$ values which correspond to the growth period of the spatially larger fluid inclusion sample (Fig. S4). The calculated fractionation factor $^{18}\alpha(CaCO_3$-$H_2O)$ between calcite and fluid inclusion water yields values between 30.7 and 32.9 ‰. This range would correspond to temperatures between 3.5 ± 1.5 °C and

12.5 ± 1.5 °C using the $^{18}α(CaCO_3-H_2O)$-T relationship of Kim and O'Neil (1997) (Table 4). The calculated absolute temperatures deviate slightly from these values depending on the used $^{18}α(CaCO_3-H_2O)$-T relationship (e.g., Démeny et al., 2010; Tremaine et al., 2011). However, relative differences between the coldest and warmest periods and the trend in the data set is largely independent of the selected fractionation-temperature relationship as most experimental and empirical studies yield similar $^{18}α(CaCO_3-H_2O)$-T slopes. Following an apparent change of 2.2 ‰ in $^{18}α(CaCO_3-H_2O)$ a temperature change of about 9°C would formally correspond to the growth period of the stalagmite. This temperature difference is much larger compared to that observed at local meteorological stations with maximum and minimum mean annual air temperature differing by approximately 3°C. This discrepancy suggests that the temperature trend related to $^{18}α(CaCO_3-H_2O)$ in the stalagmite has been enhanced, e.g., by stronger isotopic disequilibrium. As the measured fluid inclusion water isotopes correspond to the meteoric water line (Fig. S5), post-depositional and other significantly altering effects are unlikely for the water-filled inclusions. However, mineral formation in speleothems often takes place in a non-equilibrium regime (Deininger et al., 2021) and may also have influenced the calcite $δ^{18}O$ values of Stam 4 due to a high growth rate and strong seasonal variations in prior calcite precipitation (PCP, Warken et al., 2018).. We refrain from correcting the disequilibrium effect in calcite $δ^{18}O$ values due to the related large and hardly quantifiable uncertainties and only focus on the fluid inclusion $δ^2H$ values in the following. Note, that it may be possible in other cases to derive temperature variations from the oxygen isotope fractionation between fluid and calcite if the degree of PCP is negligible or constant and the length of drip interval has not changed significantly during growth.

Affolter et al. (2019) demonstrated that $δ^2H$ values and its temperature relationship in rainwater of mid-latitudes can be used to deduce temperature changes throughout the Holocene. In stalagmite Stam 4, a long-term trend towards higher $δ^2H$ values is observed from the oldest to youngest fluid inclusion samples (Fig. 9). A significant increase for $δ^2H$ values of +4.8 ± 2.1 ‰ was identified between sample level F and K and similarly between B and C (Fig. 9 C). This transfers into to a temperature change of +1.0 ± 0.4 °C using the relationship between the isotopic composition of precipitation and temperature for Central Europe of +0.59 ± 0.04 ‰/°C for oxygen and +4.72 ± 0.32‰/°C for hydrogen isotopes (Rozanski et al., 1992). Since stalagmite Stam 4 from Cloşani Cave grew under continental climatic influence, the mean value for Central Europe seems to be the best reference for the determination of the relative temperature change with the $δ^2H$/T relationship. GNIP (Global Network of Isotopes in Precipitation) stations and other weather stations in Hungary, Austria, Slovakia and Poland with more than 10 years of isotope analysis show similar slopes of +3.9-5.4 ‰/°C (Demény et al., 2021). Considering the observed range of the rainfall $δ^2H$-temperature slopes in Central and Eastern Europe by Gaussian error propagation, the uncertainty increases slightly to 0.5°C.

With the confirmed recent growth of the stalagmite, the topmost stalagmite piece is assigned to the year 2010 C.E.( year of stalagmite removal). Annual growth layers provide a possibility to assign ages to all other sample depths (Supplementary Fig. S2). Temperature changes Δ$T$ relative to the reference level B is close to zero up to ca. 1960 C.E. (3 cm distance from top, level F), followed by an increase of 1.0 ± 0.4 °C at the stalagmite top (Fig. 9F). The mean annual air temperature for the time period from 1928 to 2008 C.E. at the meteorological station Drobeta/Turnu Severin, which is located in the vicinity of the cave, shows a similar temperature increase of about 1 °C from 1980 until 2008 C.E. (Fig. 9G). This is consistent with the general trend in Romania, which experienced a 0.8 °C increase for the period of 1901-2012 C.E. (Ministry of Environment and Climate Change, 2013). The temperature change determined from $δ^2H$ values in the fluid inclusions corresponds well to the trend and magnitude measured in the mean annual air temperature of the region (Fig. 9). Directly interpreting the fluid inclusion $δ^{18}O$ values using the rainfall $δ^{18}O$-T relationship for Central Europe of +0.59 ± 0.04 ‰/°C by Rozanski et al. (1992) also leads to a temperature increase, albeit with a higher amplitude of 2.0 ± 1.1°C relative to level B, but within uncertainty consistent with the temperature reconstruction using $δ^2H$ values (Table 4).

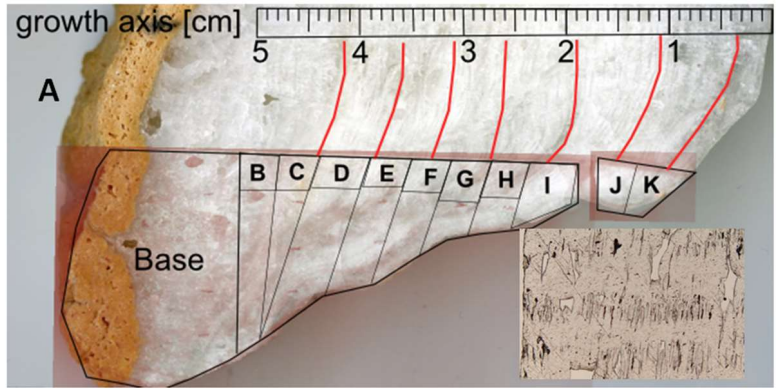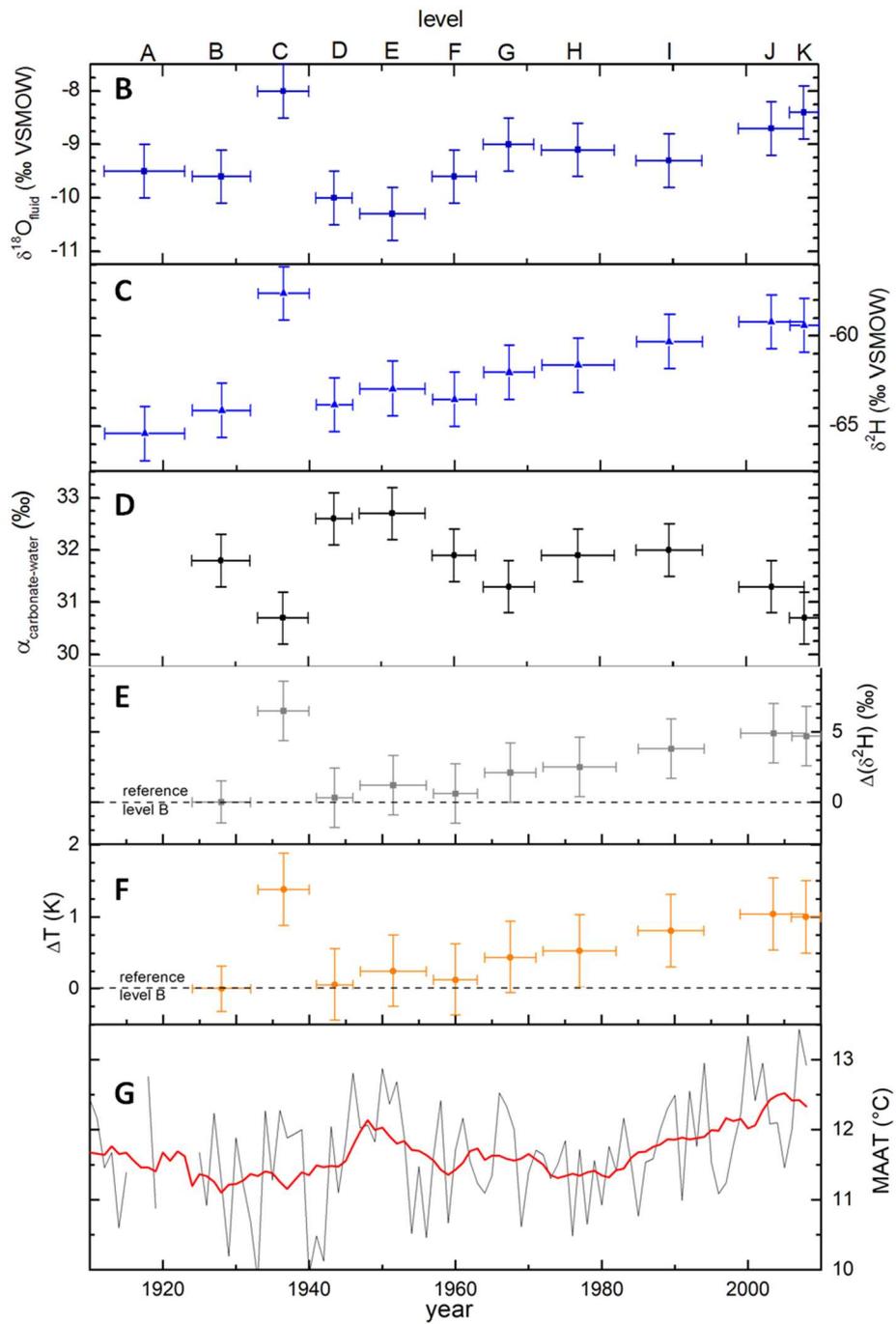

Figure 9: A) Stam 4 with assignment of sample pieces based on visual correlation with the growth axis. Layers B to K were used for temperature reconstruction. The small inset shows alternation of fluid inclusion-rich and inclusion-poor layers. Winter layers yield very little inclusions, while summer layers include abundant air- and water-filled inclusions. Width of the image is ca. 3 mm. B) Fluid inclusion $\delta^{18}O$ values C) $\delta^2H$ values D) Fractionation factor α between calcite and inclusion water. E) Change in $\delta^2H$ values relative to level B (lowest temperature). F) Inferred temperature change relative to level B. A trend is visible for Δ($\delta^2H$) as well as for ΔT from the stalagmite bottom to the top. Using the $\delta^2H$/T relationship of 4.72 ± 0.32 ‰/T (*Rozanski et al.*, 1992) a total increase of ΔT =1.0 ± 0.5 °C is observed within the growth period of the investigated stalagmite. G) Mean annual air temperature (MAAT) of the Drobeta/Turnu Severin station in the cave region (thin black line, *Klein Tank et al.*, 2002) for the last 100 years with a 10-year running mean (red line) . The 10 year-smoothing interval corresponds to the average age that is covered by the fluid inclusion samples. The uncertainties are given on the 1σ level. [black/white for figures in print, colour online]

## *4.3 Paleotemperature reconstruction using fluid inclusions*

Our study supports the conclusion of previous publications (e.g., Affolter et al., 2014; Uemura et al., 2016; de Graaf et al., 2020) that an accurate and precise determination of the isotope composition of micro-litre water amounts is possible. Our setup is able to produce small errors, which are in the same range as the precision in the previous fluid inclusion isotope studies(Dublyansky and Spötl,2009; Arienzo et al., 2013); Affolter et al., 2014),; Uemura et al. 2016; Dassié et al.,2018). In these studies a precision of 0.3-0.5 ‰ for $\delta^{18}O$ and 0.7-1.9 ‰ for $\delta^2H$ values in the water amount range of 0.1-1.0 μl, and 0.1-0.3 ‰ for $\delta^{18}O$ and 0.2-0.7 ‰ for $\delta^2H$ values at water amounts > 1 μl was demonstrated. The analytical precision determines the currently achievable temperature precision.

In principle, three possible ways of temperature calculation from fluid inclusion isotopes exist: a) from the temperature-dependent oxygen isotope fractionation between calcite and fluid inclusion water (e.g., Arienzo et al., 2015; Labuhn et al., 2015), b) indirectly via transfer of the fluid inclusion $\delta^2H$ value to the corresponding water $\delta^{18}O$ value using the $\delta^{18}O$-$\delta^2H$ relationship of the meteoric water line and then using the oxygen isotope fractionation between carbonate and water for temperature calculation (Zhang et al., 2008; Meckler et al., 2015), and c) from the hydrogen isotopes using a locally valid $\delta^2H$-temperature relationship of the rainfall (e.g., Affolter et al., 2019). Of the three methods for temperature reconstruction the first two (a and b) show the highest uncertainty of 0.6-3.1 °C (Van Breukelen et al., 2008; Zhang et al., 2008; Arienzo et al., 2015; Meckler et al., 2015; Uemura et al., 2016). The highest achievable temperature precision in the case of the best analytical fluid inclusion $\delta^{18}O$ precision of 0.1-0.2 ‰ (and a calcite $\delta^{18}O$ uncertainty <0.1 ‰) is 0.6-1.1 °C. Approaches a and b are additionally affected by the potential influence of disequilibrium isotope fractionation during carbonate mineral formation (e.g., Deininger et al., 2021), causing too high temperatures or an unrealistically large temperature spread in case of significant changes of isotopic disequilibrium.

Furthermore, diagenetic exchange between host calcite and fluid inclusion water could further alter the water $\delta^{18}O$ value (Demeny et al., 2016; Uemura et al., 2020). The precision of the temperature reconstruction directly from fluid inclusion $\delta^2H$ values depends critically on the value of the rainfall $\delta^2H/T$ relationship and the availability of well-defined rainfall $\delta^2H/T$ functions at the study site. For the Central European region a value of 4.72 ± 0.32 ‰/°C of Rozanski et al. (1992) can be used and can yield a temperature precision of 0.2 °C for released water amounts of ~0.5 µl if the analytical precision is ~1.0 ‰ for $\delta^2H$ measurements. The uncertainty of the rainfall $\delta^2H/T$ function is negligible for our case study but could be relevant in case of a reduced temperature dependence of the rainfall $\delta^2H$ values. At locations with a stronger temperature dependence of the rainfall $\delta^2H$ value an even better precision is possible, e.g., ± 0.13 °C for the average of Swiss stations, which show a slope of 7.44 ‰/°C (Rozanski et al., 1992) and for the typical analytical uncertainty of our setup.

The temperature resolution of this method is slightly reduced at lower latitudes (e.g., ± 0.55 °C at Hong Kong with a $\delta^2H$ rainfall-temperature relationship of 2 ‰/°C; Rozanski et al., 1992). Note that temperature estimates using fluid inclusion $\delta^2H$ values and the rainfall $\delta^2H/T$ relationship without climatic reference points are relative, i.e., they record only temperature changes. With an anchor, e.g., modern reference temperature and rainfall $\delta^2H$ values, absolute temperatures can be also inferred from fluid inclusion $\delta^2H$ values. The application of the rainfall $\delta^2H/T$ relationship for calculating temperature changes from fluid inclusion $\delta^2H$ values also requires the $\delta^2H/T$ relationship to be constrained for the past. Information on the rainfall isotope systematic and the $\delta^2H/T$ relationship in the past can be gained for example from groundwater studies (Darling, 2004) in combination with noble gas temperatures (e.g., Kreuzer et al., 2009; Varsány et al., 2011, Túri et al., 2020). The uncertainty of the $\delta^2H/T$ relationship needs to be considered and likely decreases the achievable precision for pre-Holocene speleothems as the uncertainty for the $\delta^2H/T$ relationship increases when applying the modern or Holocene relationship back in time.

Affolter et al. (2019) used the $\delta^2H/T$ relationship for temperature reconstruction from fluid inclusions throughout the Holocene and achieved a precision of 0.2-0.5 °C for a Swiss stalagmite. Our analytical approach allows for the same temperature resolution and with measurements of stalagmite Stam 4 from Romania confidently verified the recent 20th century warming. Both studies together illustrate the potential of the inclusion-based methodology for tracing and reconstructing minor temperature fluctuations of < 2 °C during the Holocene and, at sufficient temporal resolution (requiring high stalagmite growth rates), also of sub-degree changes such as the recent anthropogenic warming trend.

## ***5. Summary and conclusion***

Fluid inclusion isotope analysis using CRDS measurements after mechanical sample crushing benefits from fluid extraction and measurement under a constant and controlled water vapour background. The specific isotope and water volume calibration of the CRDS system remained valid for several years. For assessing the fluid inclusion extraction and measurement performance we used syringe injections and boro-silicate glass capillaries filled with reference water. We have shown that out setup has no drift in the isotope values for smaller water amounts and that the memory effect for this system is negligible when using an isotopically appropriate background water vapour. The water vapour background should be chosen such that the isotope values of sample and background do not deviate significantly (maximum 10 ‰ for $\delta^{18}O$ and 50 ‰ for $\delta^{2}H$ values).

Direct comparison of calcite powder-free and -filled extraction tubes proved that the adsorption of water on the speleothem surface has no effect on the measured isotope signal if the water content is larger than 1 µl water per g calcite. For samples with a water content below 0.1 µl/g calcite results have to be checked as we observed a corresponding adsorption of the water vapour background on freshly crushed calcite. Related to the above-mentioned constraints, the precision (1σ) of isotope measurements for aliquots of water from speleothem fluid inclusions improves with increasing water amount. It is 0.4-0.5 ‰ for $\delta^{18}O$ and 1.1-1.9 ‰ for $\delta^{2}H$ values for water samples between 0.1 and 0.5 µl, which is comparable to other CRDS systems and IRMS techniques. This value was further confirmed by replicated measurements of adjacent samples of the Romanian stalagmite Stam 4 (standard deviation of 0.5 and 1.2 ‰ for $\delta^{18}O$ and $\delta^{2}H$ values). For water amounts larger than 1 µl the precision improves to 0.1-0.3 ‰ for $\delta^{18}O$ and 0.2-0.7 ‰ for $\delta^{2}H$.

Analysis of fluid inclusions of recent pool spars from a German cave shows good agreement between drip water and fluid inclusion isotope values. Similarly, the $\delta^{18}O$ and $\delta^{2}H$ values of a Romanian stalagmite, grown during the 20th century, reflect the isotopic composition of the modern drip water within uncertainty. In the same case study, we observed a T-trend from $\delta^{18}O$ values, which is inconsistent with local weather records, suggesting a major influence disequilibrium and kinetic effects on the speleothem calcite $\delta^{18}O$ signal of Stam 4. The isotopic disequilibrium causes a significant overestimation of the temperature changes calculated from the oxygen isotope fractionation between calcite and water (in our case 9 °C difference instead of ca. 1°C). In contrast, hydrogen isotopes are not involved in calcite precipitation and therefore provide a relatively undisturbed link to the stable isotopic composition of drip and rain water. Using the $\delta^{2}H$-temperature relationship in rainfall we obtained a temperature increase for Cloşani Cave of +1.0 ± 0.5 °C between 1960 and 2010, which is in excellent agreement with the local temperature record. Thus, applying the local rainfall $\delta^{2}H$--temperature relation on fluid inclusion $\delta^{2}H$ variations appears to be a reliable method to determine mean annual air temperatures for mid-latitude speleothems. The achieved precision furthermore highlights the potential of fluid inclusion isotope studies in speleothems for high resolution paleoclimate reconstruction, given that the rainfall isotope relationship is significantly linked to temperature and is available for the studied area and valid for past periods.

**Acknowledgements**

The project was funded by DFG Grant KL 2391/2-1 and supported by the Heidelberg Graduate School for Physics in the context of grant GSC 129. We thank Sylvia Riechelmann and Jasper Wassenburg for collection of Stam 4, Silviu Constantin and Mihai Terente for monitoring, Christoph Spötl for drip water analysis at CL3, Regina Mertz-Kraus for LA-ICP-MS element analysis and Sven Brömme for calcite $\delta^{18}$O and $\delta^{13}$C analysis on Stam 4. We thank the editor Michael E. Böttcher and three anonymous reviewers for their very detailed comments and suggestions that helped to improve the manuscript.

**Data availability**

Data of this study are summarized in Tables 1-4 and Supplementary Table S1-S3. Raw data related to Figs. 3-6 are given in the Appendix of Weißbach (2020), available at https://doi.org/10.11588/heidok.00028559

**Competing interests statement**

The authors declare the absence of competing interests.

# Supplementary Material

for

# Constraints for precise and accurate fluid inclusion stable isotope analysis using water-vapour saturated CRDS techniques

by

Therese Weissbach, Tobias Kluge, Stéphane Affolter, Markus C. Leuenberger, Hubert Vonhof, Dana F.C. Riechelmann, Jens Fohlmeister, Marie-Christin Juhl, Benedikt Hemmer, Yao Wu, Sophie Warken, Martina Schmidt, Norbert Frank, Werner Aeschbach

## S1) Hüttenbläserschacht Cave – sample dating

**Table S1:** Pieces of speleothem samples (pool spars and rafts) collected at Hüttenbläserschacht Cave were dated using the Th-U disequilibrium method at the Heidelberg Academy of Sciences. The analytical procedure followed the methods described in Fohlmeister et al. (2012).

| Sample | $^{232}Th$ [ppb] | $^{238}U$ [ppb] | $^{230}Th$ [fg/g] | ± | $(^{234}U/^{238}U)$ | ± | $(^{230}Th/^{238}U)$ | ± | Age Corr. | Age uncor. |
|---|---|---|---|---|---|---|---|---|---|---|
| Hinterm Ballsaal | 13.0 | 382.1 | 58.8 | 6.7 | 1.1407 | 0.0030 | 0.0005 | 0.0011 | 0.04±0.26 | 0.90±0.08 |
| Kristall-häutchen | 3.1 | 321.9 | 34.3 | 2.5 | 1.2159 | 0.0053 | 0.0040 | 0.0005 | 0.3557±0.13 | 0.5862±0.04 |

## S2) Closani Cave and Stam 4 annual layer counting

Cloşani Cave is located on the southern slope of the Carpathians at an altitude of 433 m above sea level and developed in massive limestones of Upper Jurassic-Aptian age (*Constantin and Lauritzen*, 1999). The cave is overlain by about 30 m of rock overburden. A monitoring programme showed microclimatic stability for the cave interior with a mean air temperature of 11.4 ± 0.5 °C and a relative humidity close to 100% for 2010-2012 and 2015 (*Warken et al.*, 2018). The cave air $pCO_2$ pattern follows a strong seasonal cycle with high values in late summer (up to 8000 ppmV) and lower values during winter (2000 ppmV). Water infiltration occurs predominantly during winter time (October - March) where 75 to 100% of the meteoric precipitation is available for infiltration. *Warken et al.* (2018) showed that calcite precipitation is favoured during winter time and reduced in summer, as a result of seasonally varying $CO_2$ concentrations in the cave air and related equilibrium DIC concentrations. The water isotopic composition of the drip water in direct vicinity (1 m) of the former location of Stam 4 shows no seasonal cycle and is constant with a mean value of −9.6 ± 0.2‰ for $\delta^{18}O$ and −66.3 ± 1.7‰ for $\delta^2H$.

The relatively small and fast-grown stalagmite Stam 4 was collected from the "laboratory passage" in the cave in 2010. It has a total length of 6 cm and an average growth rate of 510 µm per year, as deduced from counting of elemental layers. Both summer and winter layers are clearly detectable in the thin sections, whereas winter layers show a compact structure with a lower number of inclusions and the milky-white porous summer layers contain abundant air- and water-filled inclusions. This layering in Stam 4 was induced by the strongly changing $pCO_2$ in the cave air, resulting in a seasonal change in growth rate and corresponding seasonal cycles in Sr and Ba in the stalagmite calcite. Similar seasonal Sr and Ba pattern have also been observed e.g., by Treble et al. (2003), Mattey et al. (2010), and Warken et al. (2018). The visible annual layers in stalagmite Stam 4 are not as pronounced as the annual cycles in the measured high-resolution Ba concentration. Ba concentration was measured with a LA-ICP-MS (Agilent 7500 ce with Laser UP-213, Institute of Geosciences Mainz) at 4.3 µm resolution. The minima of this record were counted five times. These five counted layer series were cross-dated to each other. Layers have to be counted at minimum three times, layers only counted once or twice were deleted from the time series. For each layer a mean value of layer thickness was calculated from the five layer thickness series to a master chronology. This layer thickness chronology results in a growth of Stam 4 from 1910 to 2010, the year of sampling under an active drip site.

**S3) Radiocarbon dating**

Four samples were drilled with a hand-held dental burr (1 mm). Calcite powder was acidified in vacuum with HCl. The emerging $CO_2$ was combusted to C with $H_2$ and an iron catalyst at 575°C (Fohlmeister et al., 2011). Measurements were performed with a MICADAS AMS system (Synal et al., 2007) in the Klaus-Tschira laboratory Mannheim. The results for the four samples show a typical speleothem radiocarbon bomb spike (Tab. S1, Fig. S1), constraining recent growth of the speleothem.

*Table S2*: *Radiocarbon measurement results. Radiocarbon results and errors are expressed in fraction modern (fm).*

| MAMS lab nr. | depth [mm] | $^{14}C$ [fm] | $^{14}C$ error [fm] |
|---|---|---|---|
| 14709 | 0.5 | 1.0416 | 0.0029 |
| 14710 | 18.7 | 1.0730 | 0.0029 |
| 14711 | 39 | 0.9265 | 0.0025 |
| 14712 | 55 | 0.9133 | 0.0024 |

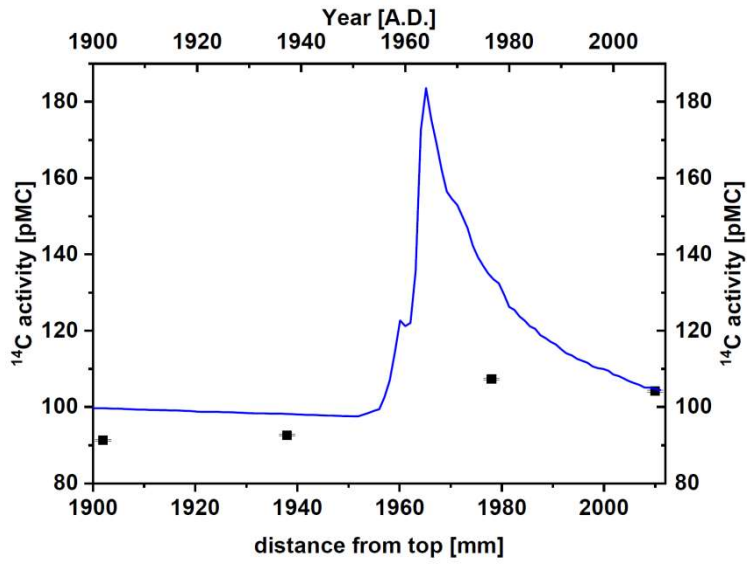

**Fig. S1**: Radiocarbon measurements (black) over depth (bottom-axis), plotted to fit the atmospheric radiocarbon anomaly (blue, top x-axis) in the mid to late 20[th] century.

# Additional figures

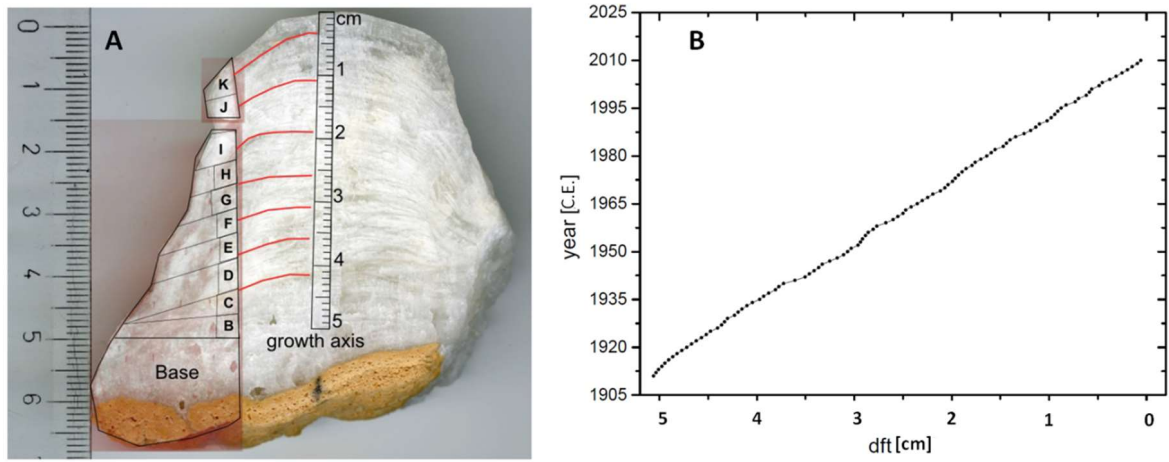

**Fig. S2**: **Age assignment of the fluid inclusion samples**

A) Fluid inclusion sample pieces (labelled B to K) are shown on the left half of the stalagmite slab. The red lines illustrate the assignment of the individual sample blocks to the growth axis. The visible lamination was used as guideline for correlation. Sample A is related to the base and due to a disturbed growth structure does not allow to assign any age. Due to the intrinsic uncertainties of this procedure (for details see Weißbach, 2020) we associated age ranges to the individual fluid inclusion samples B to K.

B) Age depth model with distance from top (dft) in cm. The chronology was established by layer counting and additional $^{14}$C measurements (see S1 and S2).

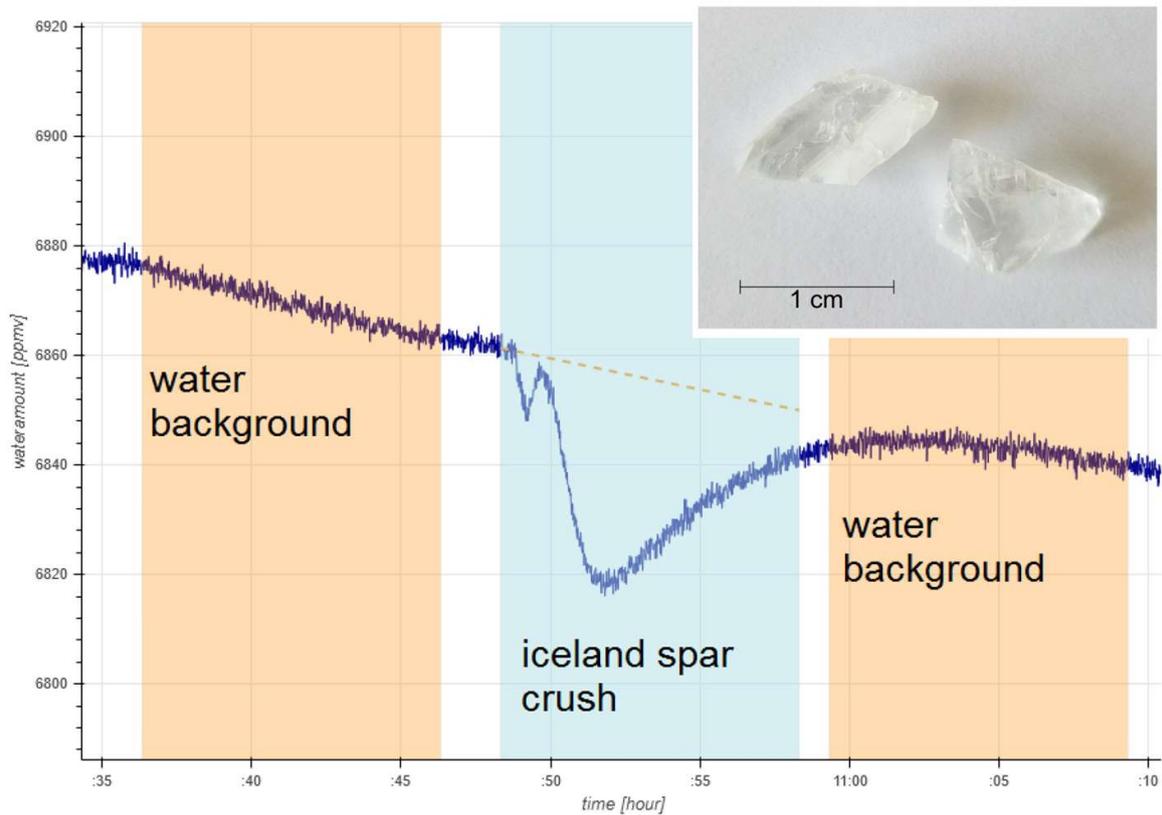

**Fig. S3**: Water vapour adsorption by the artificial fluid inclusion system. Water vapour concentration during crushing of 0.25 g Iceland spar. The decrease of the water vapour concentration indicates an adsorption of water molecules on the freshly crushed calcite. Using the water amount calibration, it corresponds to about 0.023 µl of water adsorption. The reference water vapour background is marked in orange with interpolated linear fit as dashed line. The small inset shows examples of compact and inclusion-free pieces of Iceland spar.

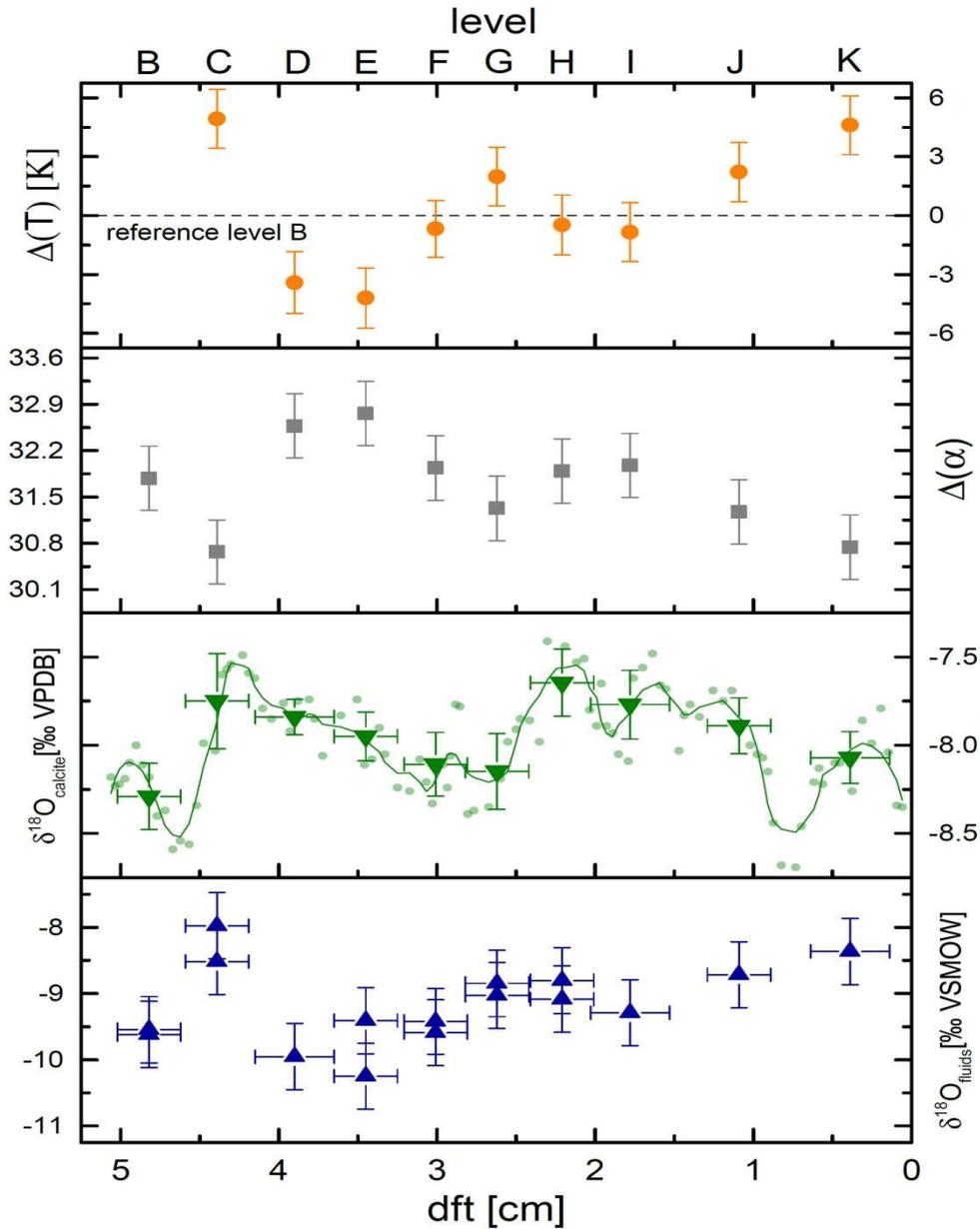

**Fig. S4**: from top to bottom: relative temperature change derived from α(CaCO$_3$-H$_2$O) relative to sample level B (orange dots); fractionation factor α(CaCO$_3$-H$_2$O) (grey squares); calcite δ$^{18}$O values (green triangles) corresponding to intervals with an edge length of 0.5 cm of the fluid inclusion sample pieces, with smoothed higher-resolution data (green line); fluid inclusion δ$^{18}$O (blue triangles). For a better overview the depth (dft) errors of α(CaCO$_3$-H$_2$O) and the calculated temperature change are not shown, but are the same as for the fluid inclusions δ$^{18}$O.

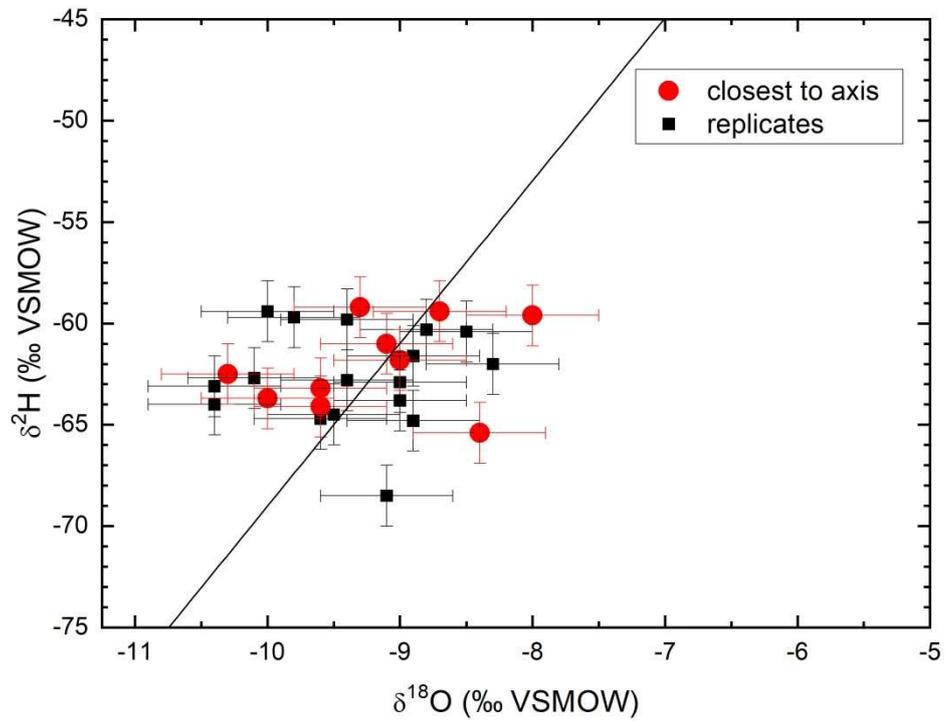

**Fig. S5**: Samples B-K of stalagmite Stam 4 with replicates from the same growth phases (Table 4) displayed relative to the meteoric water line. The aliquots closest to the growth axis of the stalagmite are shown as red circles.

## Additional Tables

**Table S3:** Fluid inclusion data from the outermost layer of stalagmite Stam 4. The distance to the growth axis increases with higher Roman numbers. Arabic numbers indicate replicates with similar distance from the growth axis. Samples in grey are not included in the interpretation and discussion as the water amount was below 0.2 µl.

| ID | Sample weight (g) | Water (µl) | Water content (µl/g) | δ²H (‰ VSMOW) | δ¹⁸O (‰ VSMOW) |
|---|---|---|---|---|---|
| I-1 | 0.69 | 0.30 | 0.52 | -65.4 ± 1.5 | -9.5 ± 0.5 |
| II-1 | 0.61 | 0.40 | 0.95 | -64.1 ± 1.5 | -9.6 ± 0.5 |
| II-2 | 0.30 | 0.37 | 0.75 | -64.7 ± 1.5 | -9.6 ± 0.5 |
| II-3 | 0.37 | 0.40 | 0.76 | -64.8 ± 1.5 | -8.9 ± 0.5 |
| II-4 | 0.38 | 0.29 | 0.56 | -68.5 ± 1.5 | -9.1 ± 0.5 |
| III-1 | 0.38 | 0.81 | 1.66 | -59.6 ± 1.5 | -8.0 ± 0.5 |
| III-2 | 0.40 | 0.51 | 1.21 | -60.4 ± 1.5 | -8.5 ± 0.5 |
| III-3 | 0.44 | 0.44 | 0.81 | -63.8 ± 1.5 | -9.0 ± 0.5 |
| IV-1 | 0.41 | 0.18 | 0.57 | -63.8 ± 1.5 | -8.5 ± 0.5 |
| IV-2 | 0.47 | 0.42 | 0.83 | -63.7 ± 1.5 | -10.0 ± 0.5 |
| V-1 | 0.23 | 0.38 | 0.78 | -62.5 ± 1.5 | -10.3 ± 0.5 |
| V-2 | 0.58 | 0.46 | 0.78 | -62.8 ± 1.5 | -9.4 ± 0.5 |
| VI | 0.47 | 0.35 | 0.75 | -63.2 ± 1.5 | -9.6 ± 0.5 |
| VII | 0.46 | 0.43 | 0.94 | -65.4 ± 1.5 | -8.4 ± 0.5 |

**Table S4:** Precision of fluid inclusion $\delta^{18}O$ and $\delta^{2}H$ measurements, interpolated from repeated water injections and crushing of water-filled glass capillaries. The values refer to an exponential fit to the standard deviation at various water amounts (Fig.5). The precision at 0.02-0.1 µl are extrapolated using the exponential fit.

| Water amount (µl) | Precision (1σ) $\delta^{18}O$ (‰) | $\delta^{2}H$ (‰) |
| --- | --- | --- |
| 0.02 | 0.55 | 2.08 |
| 0.05 | 0.54 | 2.00 |
| 0.08 | 0.53 | 1.92 |
| 0.1 | 0.53 | 1.87 |
| 0.2 | 0.50 | 1.65 |
| 0.3 | 0.47 | 1.45 |
| 0.4 | 0.44 | 1.28 |
| 0.5 | 0.42 | 1.14 |
| 0.6 | 0.40 | 1.01 |
| 0.7 | 0.37 | 0.90 |
| 0.8 | 0.35 | 0.81 |
| 0.9 | 0.34 | 0.73 |
| 1.0 | 0.32 | 0.66 |
| 2.0 | 0.20 | 0.33 |
| 3.0 | 0.14 | 0.26 |
| 4.0 | 0.11 | 0.24 |

## Additional references: